\documentclass[11pt,nofootinbib,aps,onecolumn,letterpaper,preprintnumbers,superscriptaddress,showkeys]{revtex4}
\usepackage{graphicx}
\usepackage{bm}
\usepackage{color}
\usepackage{amsmath}
\usepackage{amsfonts}
\usepackage{amssymb}
\usepackage{natbib}
\usepackage{latexsym}
\usepackage{mathrsfs}
\usepackage[usenames,dvipsnames,svgnames]{xcolor}  %% colored text
\usepackage{hyperref}   %% Needs to make hyper reference of any \definecolor{oxfordblue}{rgb}{0.0, 0.13, 0.28}
\definecolor{burgundy}{rgb}{0.5, 0.0, 0.13}
\definecolor{darkolivegreen}{rgb}{0.33, 0.42, 0.18}
\definecolor{deepmagenta}{rgb}{0.8, 0.0, 0.8}
\definecolor{richcarmine}{rgb}{0.84, 0.0, 0.25}
\definecolor{bluer}{rgb}{0.00,0.50,0.75}{}
\definecolor{darkpastelgreen}{rgb}{0.01, 0.75, 0.24}
\hypersetup{colorlinks=true, citecolor=green, linkcolor=deepmagenta,
urlcolor = cyan, filecolor=cyan}

\begin{document}
%\date{\today}
\newcommand\be{\begin{equation}}
\newcommand\ee{\end{equation}}
\newcommand\bea{\begin{eqnarray}}
\newcommand\eea{\end{eqnarray}}
\newcommand\bseq{\begin{subequations}} %solo con amsmath
\newcommand\eseq{\end{subequations}}
\newcommand\bcas{\begin{cases}}
\newcommand\ecas{\end{cases}}
\newcommand{\p}{\partial}
\newcommand{\f}{\frac}
\newcommand{\red}{\textcolor{red}}

\title{Magnetic Reconnection and Energy Extraction from a Spinning Black Hole with Broken Lorentz Symmetry}

\author {\textbf{Mohsen Khodadi}}
\email{m.khodadi@ipm.ir}
\affiliation{School of Astronomy, Institute for Research in Fundamental Sciences (IPM)\\
P. O. Box 19395-5531, Tehran, Iran}

 %---------------------------------------

\begin{abstract}
In the Penrose process and the Blandford-Znajek mechanism, the rotational energy of a black hole (BH) is extracted via particle fission and magnetic tension, respectively. Recently, inspired by a fundamental trait in plasma astrophysics known as magnetic reconnection (MR),  a new energy extraction mechanism based on the fast reconnection of the magnetic field lines inside the ergosphere has been proposed by Comisso and Asenjo. In this paper, we investigate energy extraction caused by MR in the ergosphere of a rapidly spinning BH with broken  Lorentz symmetry by a background bumblebee vector field. The desired rotating BH solution differentiates from the standard Kerr BH via the Lorentz symmetry breaking (LSB) parameter $l$, which comes from nonminimal coupling between the bumblebee field with nonzero vacuum expectation value and gravity. We find that incorporating $l<0$ in the background is in the interest of the energy extraction via MR for the fast-spinning BH surrounded by the plasma with weak magnetization, below what is expected from the scenario by Comisso and Asenjo.
Our analysis robustly indicates that the power of energy extraction and efficiency of the plasma energization process through fast MR is more efficient than the Comisso-Asenjo solution, provided that the LSB parameter is negative, $l<0$.
Compared to the Blandford-Znajek mechanism arising from the underlying background, we also show that MR is a more efficient energy extraction mechanism if $l<0$. 
\end{abstract}
%\pacs {04.70.-s, 04.50.Kd, 52.27.Ny}
\maketitle
\section{Introduction}
For years, black holes (BHs) have received much attention in astrophysics; in particular, after
the release of the Event Horizon Telescope (EHT) image \cite{EventHorizonTelescope:2019dse}, research on BH physics has entered a new golden age. Despite the intrinsic nature of BHs, which seems to be invisible, these mysterious objects are proving useful for understanding some astrophysical phenomena. 
It is well known that the energy extraction from a spinning BH is key to explaining some of the most energetic astrophysical phenomena, like relativistic jet overflows from active galactic nuclei (AGN) and gamma-ray bursts (GRBs).
One of the exciting predictions of the general theory of relativity (GTR) is that a spinning BH has free energy for exploitation. Apart from the phenomenologically worth of the energy extraction from a spinning BH, it can also shed light on the fundamentals of BH physics.
It was shown for the first time by Christodoulou \cite{Christodoulou:1970wf} that for a Kerr BH with mass $M$ and spin parameter $a$, a fraction of the BH mass equal to $M_{\rm irr} =  M \sqrt{\frac{1}{2} \left( {1+\sqrt{1-a^2}} \right)} $ 
is irreducible. The irreducible mass has a direct relation with the surface area of the event horizon, $A_H =  16 \pi M_{\rm irr}^2$, which is proportional to the BH entropy  $S_{\rm BH} \propto A_H$ \cite{Bekenstein:1973ur,Hawking:1974rv,Hawking:1974sw}. 
Therefore, 
the maximum energy that can be extracted from a BH without violating the second law of thermodynamics corresponds to the rotational energy $E_{\rm rot} = \bigg( {1-\sqrt{\frac{1}{2} \left( {1+\sqrt{1-a^2}} \right)}} \bigg) M c^2$.
By setting the extreme case $a =1$, the maximum energy that can be extracted turns out to be $E_{\rm rot} \simeq 0.29 M c^2$.
As a result, there is no conflict between energy extraction from rotating BHs and the second law of BH thermodynamics.

The idea of extracting rotational energy from a BH dates back to the seminal paper of Penrose \cite{Penrose:1969pc,Penrose:1971uk}, who mentioned it for the first time 
through designing a thought experiment in which particle fission ($0 \rightarrow 1 + 2$)  happens in the ergosphere surrounding a Kerr BH.
The essence of the Penrose process relies on conservation law.  In practice, if the angular momentum of particle $1$ is opposite to the BH rotation, the energy of particle $1$, from the infinity observer's viewpoint, is negative, meaning that to satisfy conservation law, the energy of particle $2$, which escaped to infinity should be larger than that of the initial particle $0$. However, there are some limitations to the Penrose process that make it an inefficient energy extraction method in astrophysical scenarios. 
More precisely, the Penrose process is not expected to lead to the extraction of significant rotational energy from astrophysical BHs to serve as an explanation of high energy astrophysics phenomena \cite{Wald:1974kya}.
However, alternative mechanisms were proposed for extracting BH rotational energy. Two of the most important alternatives are superradiant scattering \cite{Teukolsky:1974yv} and the Blandford-Znajek process (BZ) \cite{Blandford:1977ds}. 
The latter is of considerable phenomenological importance and is thought to be the leading mechanism for powering the jets taken out of AGNs \cite{Begelman:1984mw,Hawley:2005xs,Komissarov:2007rc,Tchekhovskoy:2011zx} and GRBs \cite{Lee:1999se,Tchekhovskoy:2008gq,Komissarov:2009dn} as well. \footnote{It is worth mentioning that energy extraction arising from the BZ process belongs to the BHs surrounded by strongly magnetized accretion, while weakly magnetized accretion,  was studied earlier by Ruffini and Wilson \cite{Ruffini:1975ne}. Similar to the Penrose process, the existence of ergoregion for the BZ process is essential, too.}

\begin{figure}[!ht]
\includegraphics[scale=0.5]{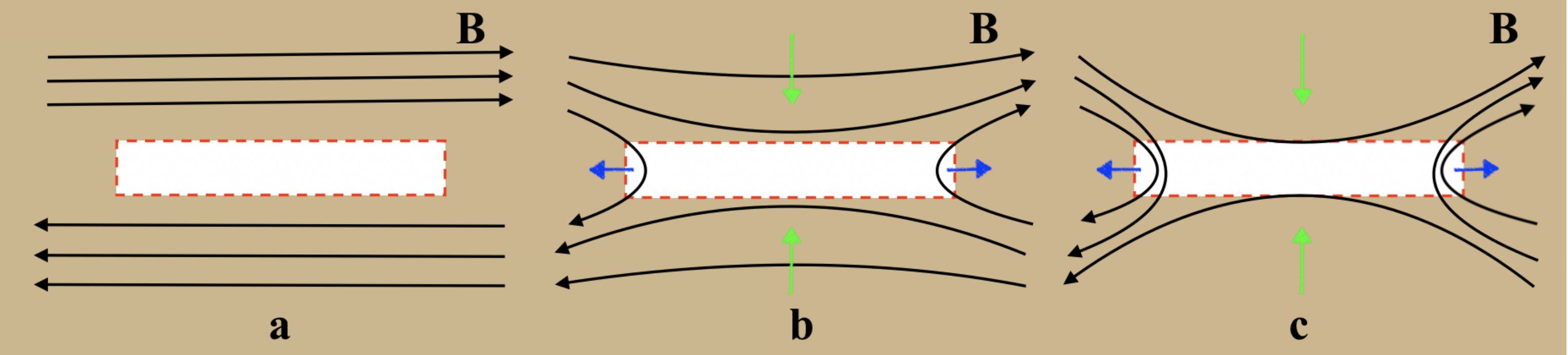}
\caption{Formation of a reconnection zone between two adjacent regions of plasma with antiparallel magnetic field lines. The region enclosed by the red dashed lines denotes the reconnection zone. The figure taken from \cite{Jafari:2018giq}.}
\label{MR}
\end{figure}

The existence of magnetic fields around BHs comes from different processes such as magnetic field generated due to matter accretion and an external magnetic field sourced by a neutron star companion near the BH. A recent analysis published by the EHT Collaboration  of polarized emission around the supermassive BH in the center of M87* \cite{Akiyama:2021tfw} also confirms the existence of a magnetic field around the BH.
One of the ideas related to a BH enclosed by the magnetic field is the magnetic reconnection (MR) mechanism. Recently, in \cite{Comisso:2020ykg}
Comisso and Asenjo proposed the idea that the rapid reconnection of magnetic field lines can be an efficient energy extraction agent. In a similar vein to the Penrose process, fast MR redistributes the angular momentum of particles efficiently to yield the negative energy at infinity, meaning energy extraction by plasma outflow from the reconnection layer towards infinity. Generally speaking, the MR usually happens in highly conducting plasma with oppositely directed magnetic field lines in which
the magnetic topology is rearranged (i.e., broken and reconnected again), and subsequently, the magnetic energy is converted to other types such as bulk kinetic energy and particle acceleration. Actually, when plasmas carrying oppositely directed magnetic field lines are brought together, a strong-current sheet is formed, known as the reconnection zone with locally strong-field know as ``reconnection zone'' with locally strong field gradients (Fig. \ref{MR}). It allows plasma to diffuse, and, thereby, magnetic reconnection will occur (see \cite{Jafari:2018giq} for more details).
	
A relevant note is that the idea of energy extraction via MR, in essence, dates back to Ref. \cite{Koide:2008xr} by Koide and Arai. 
However, that scenario suffers from several problems and does not address the energy extraction rate from the BH, or the efficiency of the energy extraction via the reconnection process. \footnote{Furthermore, unlike the scenario proposed in \cite{Comisso:2020ykg}, the reconnection process in \cite{Koide:2008xr} is not fast. The fast reconnection requires either plasmoids or turbulence. Otherwise, the magnetic diffusion regulates the reconnection speed, which is extremely low in the accretion flow. This means that the reconnection process is virtually nonexistent.} These outcomes are essential to reveal whether the MR as one of the candidates for the extraction of BH energy, is an efficient and noteworthy mechanism.
Recently, in general-relativistic kinetic simulations of BH magnetospheres \cite{Parfrey:2018dnc},  the presence of negative-energy particles has been explored, possibly associated with MR. Reference \cite{Comisso:2020ykg} envisioned a background geometry including matter and magnetic-field configuration in which fast MR occurs intermittently on the dynamical timescale; subsequently they derived the energy extraction rate from a spinning BH. This study showed that in the case of a rapidly spinning BH surrounded by a highly magnetized plasma, the reconnection process results in a considerable energy extraction rate, which can even exceed the power extracted through the BZ process.

Given the fact that all successful tests of GTR are restricted to the weak-field limit and may need some modifications in the strong-field limit, it is helpful if the GTR Kerr BH is imagined as an effective solution to a more general one. Besides, GTR fails to explain severely challenging issues such as dark matter, dark energy \cite{Clifton:2011jh}, and recent measurements of how fast the cosmos is expanding \cite{DiValentino:2021izs}. Faced with these dilemmas, part of the scientific community is relaxing some of the accepted assumptions, rules, and symmetries and is considering theories of modified gravity. 
Therefore, for a more comprehensive study of BH energy extraction via the MR,  one may want to investigate this issue for the modified gravity-based Kerr BH solutions. Such a study can be well motivated for two reasons: first, due to the dependency of the MR mechanism to the parameter(s) related to the metric background, as can be seen in \cite{Comisso:2020ykg}; second, for the appearance of the additional parameter(s) sourced by alternative models of gravity which cause the metric backgrounds to deviate from standard Kerr. It is expected that the modified metrics provide a richer phenomenology framework.

In this regard, with this paper, we intend to evaluate the feasibility conditions and efficiency of MR as an energy extraction mechanism from a rotating BH solution with broken fundamental Lorentz symmetry. Although Lorentz invariance is one of the cornerstones of modern physics, some theories propose that Lorentz invariance is energy scale dependent and, in high energy may be violated \cite{Mattingly:2005re}. Search for violation of Lorentz symmetry from a phenomenological viewpoint is well motivated because it opens an observational window on a fundamental issue as quantum gravity \cite{Kostelecky:1990pe,Kostelecky:1994rn}. The most appropriate method for implementing Lorentz symmetry breaking (LSB) into a curved space-time is based on spontaneous symmetry breaking that can occur by different mechanisms.
Bumblebee gravity (BG) \cite{Bluhm:2004ep,Bertolami:2005bh},
Einstein-Æther theory \cite{Jacobson:2000xp}, and Horava-Lifshitz \cite{Horava:2009uw} are the most well-known Lorentz invariance violation (LIV) gravity models. Concerning the BG, due to the nonzero vacuum condensation of a vector field, the so-called bumblebee field,  the LSB mechanism happens, indicating a preferred frame
\cite{Kostelecky:2003fs}. BG belongs to those modified gravity theories that break a fundamental part of GTR; i.e., Lorentz symmetry deviates from the standard model of gravity \cite{Bluhm:2008rf}.
Exact Schwarzschild-like and Kerr-like BH, and also 
traversable wormhole solutions for the underlying gravity model, are found in \cite{Casana:2017jkc,Ding:2019mal,Ovgun:2018xys} respectively.
In all three types of solutions,  the violation of  Lorentz symmetry is due to a nonzero vacuum expectation value (VEV) of the bumblebee vector field coupled to the space-time curvature.  In recent years, these solutions have been used as descriptive frameworks to study various aspects of the physics of compact objects such as BHs and wormholes at a fundamental level; see \cite{Gomes:2018oyd}-\cite{Oliveira:2018oha} for instance. 

The roadmap for this paper is as follows. In Sec. \ref{MKB} we briefly present a Kerr-like BH solution with broken Lorentz symmetry driven by the bumblebee field and discuss some of the relevant quantities. In Sec. \ref{LSB} we first delineate the extraction of BH rotational energy via MR and reveal the circumstances under which such energy extraction takes place. Next, by using parameter space analysis, we probe the conditions obtained for energy extraction in the interplay with the involved parameters, particularly the location of the reconnection and LSB parameter. In Sec. \ref{power} we compute the rate of energy extraction as well as the reconnection efficiency to evaluate the role of the LSB parameter embedded in the background. Here we also compare the power extracted by MR with the power extractable via the BZ mechanism for the underlying Kerr-like BH solution. Finally, we provide results in Sec. \ref{Con}.

%%%%%%%%%%%%%%%%%%%%%%%%%%%%%%%%%%%%%%%%%%%%%%%%%%%%%%%%%%%%%%%%%
\section{modified Kerr BH solution with a background bumblebee field}\label{MKB}
Here, we give an overview of the exact Kerr-like BH solution obtained from nonmiminal coupling the background bumblebee field to gravity.
In general, for spontaneous LSB occurring in a curved space-time, one should considers the action of the metric tensor
coupled to the vector field, which looks like the bumblebee action\footnote{The essence of this action is that its purely metric sector is presented by the standard Einstein–Hilbert action, while the dynamics of the vector field is described by the Maxwell-like term, plus a potential whose minimum yields a vector implementing the LSB. Also, there are some extra terms responsible for a vector-gravity coupling.}
\cite{Bluhm:2004ep,Bertolami:2005bh}
\begin{equation}
S=\int d^{4}x\sqrt{-g}\bigg( \frac{1}{16\pi}\left(  R+\xi B^{\mu
}B^{\upsilon}R_{\mu\nu}\right)  -\frac{1}{4}B^{\mu\nu}B_{\mu\nu}-V\left(
B^{\mu}\right)  \bigg),~~~~~~(c=1=G_N) . \label{sk1}%
\end{equation}
In the action (\ref{sk1}),
the bumblebee vector field $B_\mu$ under a proper potential 
$V\left(B^{\mu}\right)=B_{\mu}B^{\mu}\pm b^{2}$ acquires a nonzero VEV ($\langle B^{\mu}\rangle=b^{\mu}$), which means a spontaneous LSB in the gravitational sector \cite{Kostelecky:2003fs, Bluhm:2004ep}. Note that
$B_{\mu\nu}$ represents the bumblebee-field strength and is defined as 	$B_{\mu\nu}=\partial_{\mu}B_{\nu}-\partial_{\nu}B_{\mu}$.
The coupling constant $\xi$ is responsible for the nonminimal gravity interaction with the background bumblebee field.

Without mentioning details, the Kerr-like BH metric derived from the BG in the Boyer-Lindquist coordinates $x^\mu=(t,r,\theta,\phi)$ takes the following form \cite{Ding:2019mal}: 
\begin{equation}
	ds^{2}=-\left(  1-\frac{2Mr}{\rho^{2}}\right)  dt^{2}-\frac{4Mr\sqrt{l+1}a%
		\sin^{2} \theta }{\rho^{2}}dtd\phi+\frac{\rho^{2}}{\Delta
	}dr^{2}+\rho^{2}d\theta^{2}+\frac{A\sin^{2}  \theta }{\rho^{2}%
	}d\varphi^{2},\label{metric}%
\end{equation}
where%
\begin{eqnarray}
&\Delta=\frac{r^{2}-2Mr}{l+1}+a^{2}M^2,\quad \quad \rho^{2}=r^{2}+\left(  l+1\right)	a^{2}M^2\cos^{2}\theta \nonumber \\
&A=\left( r^{2}+\left( l+1\right)  a^{2}M^2\right)  ^{2}-\left(
	l+1\right)  ^{2}a^{2}M^2 \Delta\sin^{2}\theta  .\label{3}%
\end{eqnarray}
At first glance, one may think that the above metric is nothing but a standard Kerr metric since it is just enough we rescale the spin parameter $a$ as $\tilde{a} \longrightarrow \sqrt{l+1}a~ (l>-1)$. However, after doing this, we see that the metric (\ref{metric}) has a soft deviation from Kerr because the Lorentz breaking parameter $l$ still appears in the final form of the metric tensor
\begin{equation}
	g_{\mu\upsilon}=\left(
	\begin{array}
		[c]{cccc}%
		-\left(  1-\frac{2Mr}{\tilde{\rho}^{2}}\right)  & 0 & 0 & -\frac{2M^2\tilde{a} r%
			\sin^{2}\theta}{\tilde{\rho}^{2}}\\
		0 & \frac{\tilde{\rho}^{2}}{\tilde{\Delta}} & 0 & 0\\
		0 & 0 & \tilde{\rho}^{2} & 0\\
		-\frac{2M^2\tilde{a}r\sin^{2}\theta}{\tilde{\rho}^{2}} & 0 & 0 & \frac{\tilde{A}\sin^{2}%
			\theta}{\tilde{\rho}^{2}}%
	\end{array}
	\right),  \label{elemen}%
\end{equation} where
\begin{equation}
	\tilde{\rho}^{2}=r^{2}+\tilde{a}^2 M^2\cos^{2}\theta, \quad \tilde{\Delta}=\frac{r^{2}-2Mr+\tilde{a}^2 M^2}{l+1},\quad
	\tilde{A}=\left( r^{2}+ \tilde{a}^{2} M^2\right)  ^{2}-\tilde{a}^{2} M^2\tilde{\Delta} \sin^{2}\theta~.
	 \label{2}%
\end{equation}
Throughout this analysis, the rotation parameter is called $\tilde{a}$ because it absorb the effect of the LSB parameter. Indeed, the imprint of $l$ is not separable from $a$.
The inner and outer boundaries of the ergosphere region of the underlying Kerr-like BH solution read as
\begin{equation}
	r_{inn}=M+M\sqrt{1-\tilde{a}^{2}  },\text{ \ \ \ \ \ \ \ \ \ }%
	r_{out}=M+M\sqrt{1-\tilde{a}^{2} \cos^{2}\theta
	}. \label{inout}%
\end{equation}
Note that the two boundaries of the ergosphere respectively coincide with the outer event horizon $r_H$ and the static limit.
Because we need to take some quantities in our analysis as the Keplerian angular velocity $\Omega_{K}$ of bulk plasma rotating around a BH, the circular photon orbit $r_{ph}$, and innermost stable circular orbit (ISCO) $r_{isco}$, we have to compute them here. Conventionally, by considering the equatorial plane $\theta=\pi/2$, these quantities  for the underlying background are written as
 \begin{equation}\label{keplerOmega}
	\Omega_K= \pm \frac{\sqrt{M}}{r^{3/2} \pm \tilde{a} \sqrt{M^3}} \, ,
\end{equation}
\begin{equation}\label{circularorbitphotonrad}
	r_{\rm ph}=2M \bigg(1+\cos\left(\frac{2}{3} \arccos(\mp \tilde{a}) \right)\bigg) \, ,
\end{equation}
and
\begin{equation}\label{rmargbsc}
	r_{\rm isco}=M\bigg(\chi_2 +3 \mp \sqrt{(3-\chi_1)(2\chi_2+\chi_1+3)} \bigg) \, ,
\end{equation}
where 
\begin{equation}\label{}
	\chi_1=1+\big(1-\frac{\tilde{a}^2}{M^2}\big)^{1/3}\bigg(\big(\frac{\tilde{a}}{M}+1\big)^{1/3}+\big(1-\frac{\tilde{a}}{M}\big)^{1/3}\bigg),\quad
\chi_2=\sqrt{\chi_1^2+\frac{3\tilde{a}^2}{M^2}}.
\end{equation} The upper and lower signs respectively address corotating and counterrotating orbits. A simple calculation of $r_{ph}$ and $r_{isco}$ for the case of fast rotation explicitly shows that if we want to restrict MR to the ergosphere region, then we have to pick up the corotating orbits. As it is clear, because the Lorentz violating parameter $l$ is absorbed by the spin parameter by spin parameter so LSB in these three quantities leaves no trace that can be distinguishable from standard Kerr. However, in what follows, we will see that the soft deviation of metric tensor (\ref{elemen}) from its standard counterpart will leave interesting and distinguishable phenomenological imprints on the energy extraction from BH.

%%%%%%%%%%%%%%%%%%%%%%%%%%%%%%%%%%%%%%%%%%%%%%%%%%%%%%
\section{Lorentz symmetry breaking and energy extraction via MR mechanism}\label{LSB}
\begin{figure}[!ht]
\includegraphics[scale=0.65]{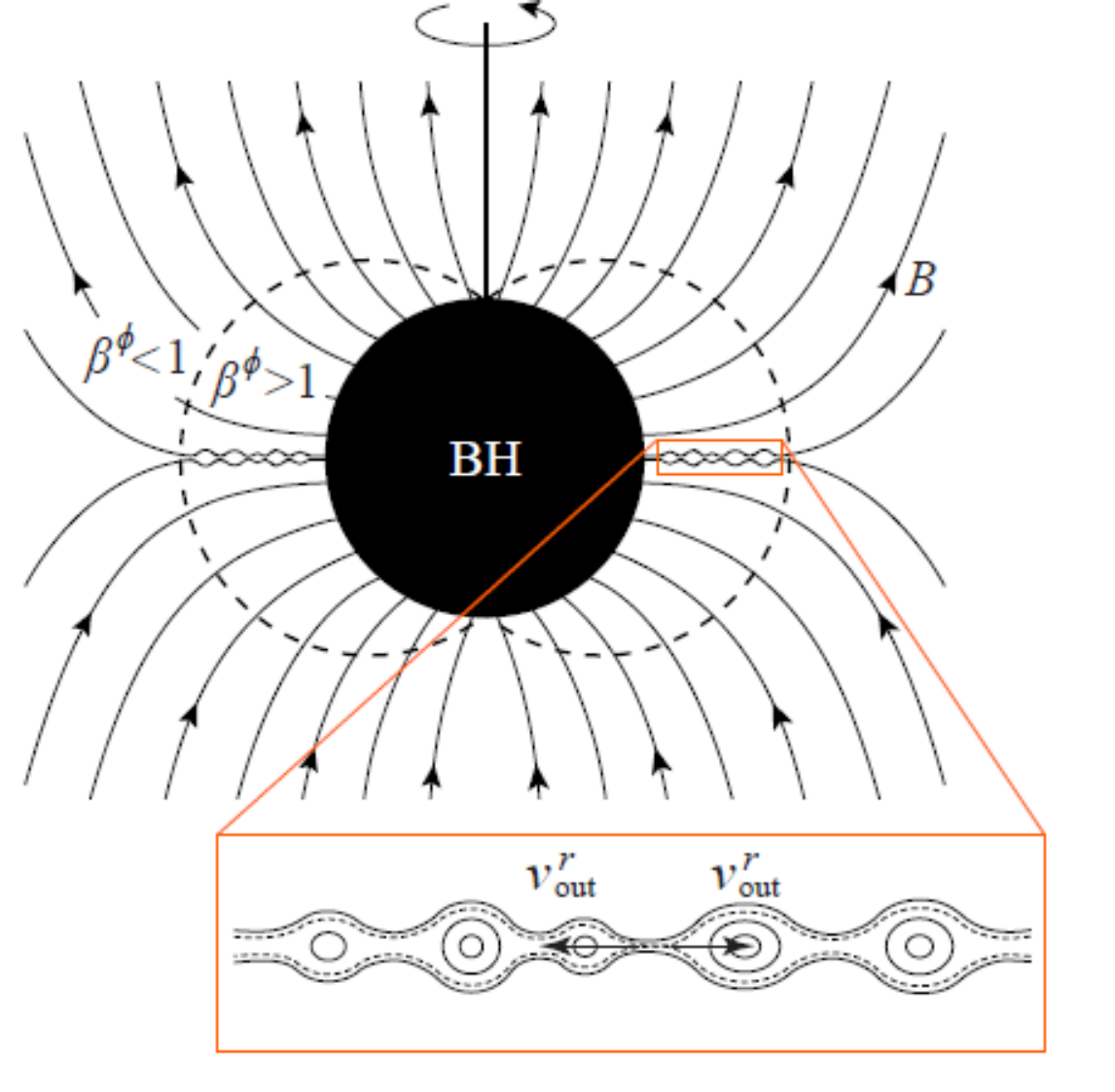}~~~~~~~
\includegraphics[scale=0.65]{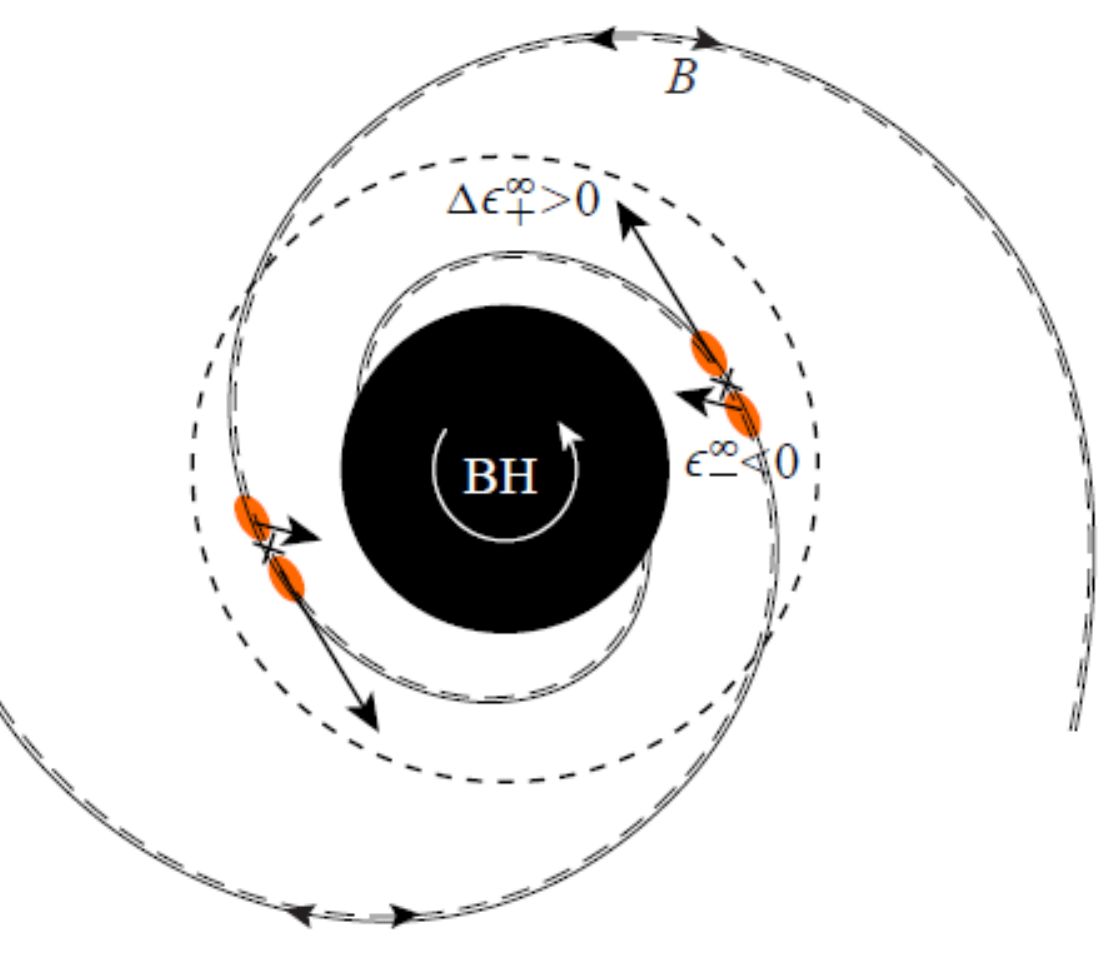}
\caption{\textbf{Left:} portrait of meridional view of a configuration with antiparallel magnetic field lines near the equatorial plane of a spinning BH. To have such a configuration, the existence of a radial component for the magnetic field lines is required.
\textbf{Right:} portrait of equatorial view of the energy extraction from a spinning BH by magnetic reconnection in the ergosphere region. Here, long-dashed and solid lines represent magnetic field lines below and above the equatorial plane, respectively. In both portraits the short-dashed lines refer to the static limit of the ergosphere. The figure taken from \cite{Comisso:2020ykg}.}
\label{bh}
\end{figure}

From a faraway observer's viewpoint, in the negative-energy orbits in regions out of even horizon and below the static limit, for extracting BH rotational energy via negative-energy particles, one needs the MR mechanism within the ergosphere region of the rotating BH.  Routinely, it is expected that the fast-spinning BHs support the MR inside the ergosphere \cite{East:2018ayf}.
In other words, the frame-dragging effect of a fast-spinning BH is naturally able to create a configuration with opposite magnetic field lines (as schematically shown in Fig. \ref{MR}) that is prone to MR. 
Therefore, we consider the configurations envisioned by Comisso and Asenjo \cite{Comisso:2020ykg}, as reported in Fig. \ref{bh}. This configuration is consistent with the numerical simulations of rapidly spinning BHs \cite{Komissarov:2005wj, Parfrey:2018dnc, Ripperda:2020bpz,Bransgrove:2021heo}. Furthermore, evidence of the existence of high-rotation BHs in the center of galaxy M87* by the EHT team \cite{Akiyama:2019fyp} is a phenomenological signal indicating that the main condition behind this idea is realistic. It is also interesting that Ref. \cite{Koide:2008xr} considered a configuration in which the current sheet within the oppositely directed magnetic field lines occurs in a plane perpendicular to the equatorial plane. However, this conflicts with the numerical simulations, e.g., Refs. \cite{Komissarov:2005wj, Parfrey:2018dnc, Ripperda:2020bpz,Bransgrove:2021heo}, which consistently show the presence of a reconnecting current sheet in the equatorial plane.

The change of direction of magnetic field lines at the equatorial plane produces a current sheet that fades against the plasmoid instability sourced by nonideal magnetohydrodynamic effects when the current sheet exceeds a critical aspect ratio \cite{Comisso:2016pyg, Comisso:2017arh}.
Indeed, here the driver of fast MR is the formation of flux ropes or plasmoids, as displayed in the zoomed-in region of Fig. \ref{bh} (left). This causes the fast conversion of the available magnetic energy into plasma particle energy, and finally, the plasma comes out of the reconnection layer \cite{Bhattacharjee:2009}. 
It has been extensively demonstrated in a large number of numerical simulations (see, e.g, Refs. \cite{Parfrey:2018dnc,Ripperda:2020bpz,Bransgrove:2021heo}) that in a reconnection layer there is always a dominant X point (reconnection location) that expels the plasma including the plasmoids or flux ropes along the direction of the neutral line, where magnetic field lines reconnect. So in MR theory and relevant simulations, it is an established fact that Plasmoids or flux ropes are particles of finite extent that move in opposite directions, as displayed in the left two-dimensional cartoon of Fig. \ref{bh}. Although a three-dimensional configuration and evolution is more complex than the two-dimensional counterpart, it is shown that the formation of flux ropes and their evacuation is similar \cite{Sironi:2014jfa}.
The field lines are then stretched again by the frame-dragging effect arising from the fast rotation of the BHs, and this trend of throwing out the plasma from the reconnection layer is repeated, as proposed by Comisso and Asenjo \footnote{It is important to note that the scenario \cite{Comisso:2020ykg} is still not fully established and awaits numerical confirmation via future studies.} \cite{Comisso:2020ykg}.
An important point to note is that based on the MR theory and simulations the particle acceleration due to the reconnection electric field accounts for only a tiny fraction of the energy gained by the accelerated particles, while most of the energy gain is due to the motional electric field \cite{Guo:2020fni}. Additionally, most of the energized particles are trapped in the plasmoids and carried out by them, as demonstrated numerically in the context of BH magnetospheres \cite{Bhattacharjee:2009}. As a result, in this scenario, the magnetic field (not the electric field) orientation near the BH horizon is the relevant one for the energy extraction from the BH. In other words, it is distinguished from the Penrose process since the principal candidate for the energy extraction is the plasma with plasmoids or flux ropes that expels along the neutral reconnection line of the magnetic field.

MR in the rotating plasma around the BH accelerates part of the plasma and decelerates another part in opposite directions. If the decelerated and accelerated parts of the plasma respectively have negative energy and energy larger than its rest mass and thermal energies at infinity, the plasma that escapes to infinity can get energy from rotation. More precisely, this happens when the negative-energy particles are swallowed by the BH. This description is schematically depicted in Fig. \ref{bh} (right). The key point is that these conditions for energy extraction can be satisfied by redistributing the angular momentum of the plasma by a fast MR process in the ergosphere of the BH. In this regard, we wish to investigate this issue for a BH background with broken Lorentz symmetry which is derived from bumblebee gravity in the above section. To constitute an efficient energy release channel, we evaluate the effect of the LSB parameter $l$ on the extraction of rotational energy from the BH in interplay with a fast MR mechanism. 

%Figure reproduced with permission from the authors in  \cite{Comisso:2020ykg} and $\textcopyright$ APS.
\subsection{Energy at infinity associated with the accelerated/decelerated plasma}

In the following, by conveniently going to the ``zero-angular-momentum-observer'' (ZAMO) frame, which is in essence a locally nonspinning frame, we compute the plasma energy density.  In the ZAMO frame, the square of the line element is given by $d{s^2} =  - d{{\hat t}^2} + \sum\nolimits_{i=1}^3 {{{(d{{\hat x}^i})}^2}}  = {\eta _{\mu \nu }}d{{\hat x}^\mu }d{{\hat x}^\nu }$, where
\begin{equation}
d\hat t = \alpha \, dt \, , \quad \; d{{\hat x}^i} = \sqrt{g_{ii}} \, d{x^i} - \alpha {\beta^i}dt \, 
\end{equation}
Note that quantities recorded in the ZAMO \footnote{
Vectors in the ZAMO frame are related to vectors with the contravariant components in the Boyer-Lindquist coordinates as $\hat b^{0}=\alpha b^{0}$ and $\hat b^{i}= \sqrt{g_{ii}} \, b^{i} - \alpha\beta^i b^{0}$, while for the covariant components, $\hat b_{0}=b_{0}/\alpha + \sum\nolimits_{i=1}^3 {(\beta^i/\sqrt{g_{ii}}) \, b_i} $ and $\hat b_i= b_i/\sqrt{g_{ii}}$.} frame are symbolized by a hat.
Here $\alpha$ and $\beta^i$  respectively denote the lapse function and the shift vector $(0, 0, \beta^\phi)$, which are defined as$ \sqrt{-g_{tt} + \frac{g_{\phi t}^2}{g_{\phi\phi}}} $ and $\frac{ g_{\phi t}\, \omega^\phi}{\alpha \sqrt{g_{\phi\phi}}}$.
By taking into account elements of the metric tensor  (\ref{elemen}) and setting the equatorial plane $\theta=\pi/2$, these read as
\begin{equation}
\alpha=  \sqrt{\frac{r^2\tilde{\Delta}}{\tilde{A}}}~,~ 
\quad \beta^\phi =  \frac{2M^2\tilde{a}}{r \sqrt{\tilde{\Delta}}}~.
\end{equation}
Now, we assess the ability of the MR in the ergosphere to extract energy from the BH via investigating the requirements for the formation of negative energy at infinity and the escaping of the accelerated or decelerated plasma to infinity. 
By adopting a one-fluid approximation for the plasma, we have 
\begin{equation}
T^{\mu \nu} = p g^{\mu \nu} +\mathscr{H} U^{\mu}  U^{\nu} + {F^\mu}_{\delta} F^{\nu \delta} - \frac{1}{4}  g^{\mu \nu} F^{\rho \delta} F_{\rho \delta} \, ,
\end{equation}
where $p$, $\mathscr{H}$, $U^{\mu}$, and $F^{\mu \nu}$ denote the proper plasma pressure, enthalpy density, four-velocity, and Farady tensor, respectively.
By defining the ``energy-at-infinity'' density $e^\infty = - \alpha g_{\mu 0} T^{\mu 0}$ which is actually total energy including
the hydrodynamic energy-at-infinity density and the electromagnetic energy-at-infinity density i.e. $e^\infty = e^\infty_{\rm hyd} + e^\infty_{\rm em}$, we have \cite{Comisso:2020ykg}
\begin{equation}\label{enerhyd}
e^\infty_{\rm hyd}  = \alpha (\mathscr{H} \hat\gamma^2 - p) + {\alpha \beta^\phi \mathscr{H} \hat\gamma^2 {\hat v}^\phi },\quad\quad  e^\infty_{\rm em}  = \frac{\alpha}{2} ({\hat B}^2 + {\hat E}^2) + \big(\hat{\textbf{B}} \times \hat{\textbf{E}}\big)_\phi~,
\end{equation}  with  $\hat\gamma =  \hat U^0 = \frac{1}{ \sqrt{1 -  \sum\nolimits_{i=1}^3 {{{(d{{\hat v}^i})}^2}}}}$,  $\hat B^i = \epsilon^{ijk} \hat F_{jk}/2$, and $\hat E^i = \eta^{ij} \hat F_{j0} = \hat F_{i0}$, which are the Lorentz factor, and the components of magnetic and electric fields, respectively. Here ${\hat v}^\phi$ denotes the azimuthal component of the outflow velocity of plasma from the ZAMO's observer viewpoint.
With the assumption that the magnetic reconnection process is efficient enough (i.e. it happens fast enough) to convert a significant part of the magnetic energy into kinetic energy, so one can neglect the contribution of $e^\infty_{\rm em}$ in total energy, which leads to 
\begin{equation}\label{enerhydincompress}
e^\infty=	e^\infty_{\rm hyd}  = \alpha  \Big(  (\hat\gamma + \beta^\phi \hat\gamma {\hat v}^\phi)\mathscr{H} -  \frac{p}{\hat\gamma}    \Big) \,  .
\end{equation}
To evaluate the reconnection process at the small scale, we have to introduce the local rest frame $x^{\mu \prime}=(x^{0 \prime}, x^{1 \prime}, x^{2 \prime}, x^{3 \prime})$ in which the directions of $x^{1 \prime}$ and $x^{3 \prime}$ are parallel to the radial direction $x^{1 \prime}=r$ and  parallel to the azimuthal direction $x^{3 \prime}=\phi$, respectively. 
For simplicity, it is assumed that from the perspective of the ZAMO observer, the bulk plasma in the equatorial plane rotates circularly with Keplerian velocity, i.e., ${\hat v}^\phi=\hat v_K$, 
\begin{equation}\label{keplerv}
\hat v_K = \frac{\tilde{A}}{r^3\tilde{\Delta}^{1/2}} {\left( {  \frac{ \tilde{a} X^{-2}-X^{-1/2}  }{\tilde{a}^2 X^{-3}-1}} \right)} -\beta^\phi \, ,
\end{equation} where $X=r/M$ is the location of reconnection, which covers range $1<X<2$ into ergosphere. By using the ``relativistic adiabatic incomprehensible ball approach'' then the hydrodynamic energy at infinity per enthalpy of the plasma thrown via the magnetic reconnection mechanism into the $\pm x^{3 \prime}$ direction, takes the following form (see \cite{Comisso:2020ykg} for more details):
\begin{eqnarray}\label{energuisMagnet}
	\epsilon^\infty_\pm \!=\dfrac{e^\infty_{\rm hyd}}{\mathscr{H}}=\! \alpha (1-\hat v_K^2)^{-1/2} \Bigg( \left(1 \!+\! \beta^\phi {\hat  v_K} \right) (1+\sigma_0)^{1/2}\pm \cos \xi ({\hat v_K}+\beta^\phi)  \sigma_0^{1/2}- \\ \nonumber
	 \frac{(1+\sigma_0)^{1/2}\mp\cos \xi \hat v_K \sigma_0^{1/2}}{4(1-\hat v_K^2)^{-1}(1+\sigma_0-\cos^2\xi {\hat v_K^2}\sigma_0)}\,  \Bigg)\, .
\end{eqnarray}
Here, $\sigma_0$, and $\xi$ respectively denote the plasma magnetization and the orientation angle, which is actually the angle between the magnetic field lines and the azimuthal direction in the equatorial plane
of the BH. As a result, the above expression for the energy at infinity associated with the accelerated (+ sign) or decelerated (- sign) plasma is, in essence, a function of the critical parameters ($\tilde{a}$, $X$, $l$, $\sigma_0$, $\xi$). The first three parameters are related to the background, while the rest come from the matter disc around the BH.

\begin{figure}
	\begin{center}
		\includegraphics[scale=0.37]{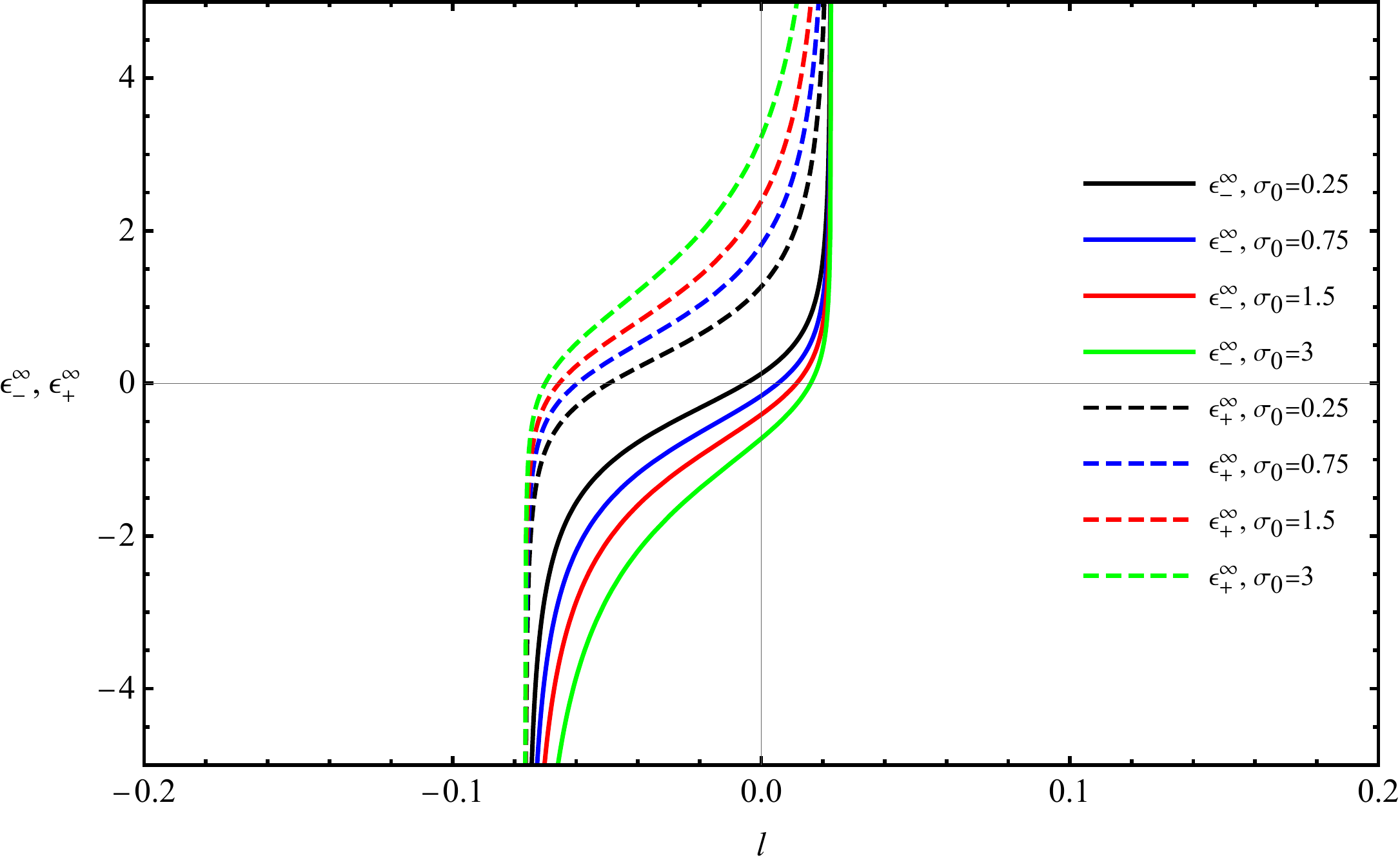}~~
		\includegraphics[scale=0.37]{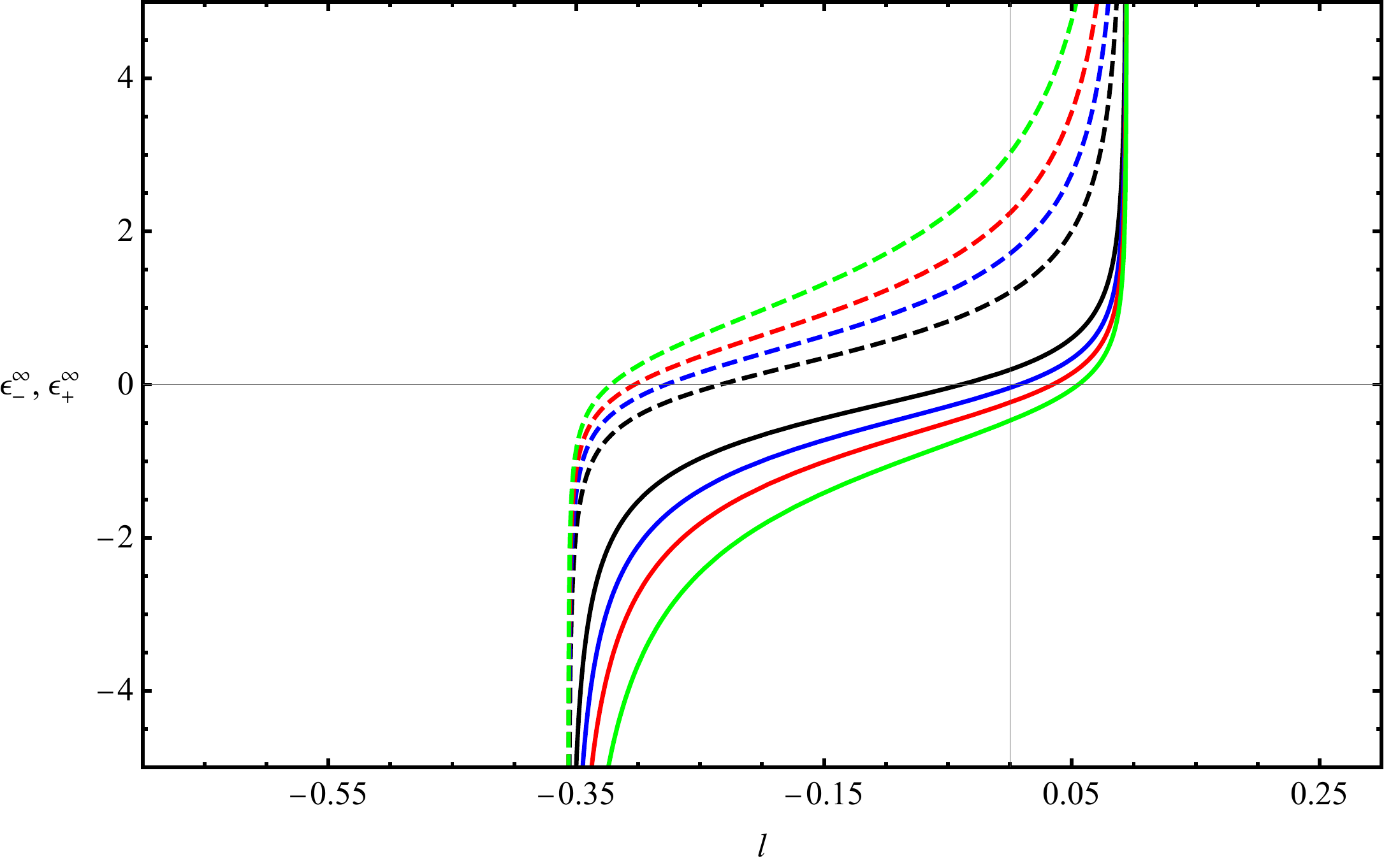}
		\caption{Behavior of $\epsilon^{\infty}_{-}$ and $\epsilon^\infty_+$ in terms of the Lorentz-violating parameter $l$ for different values of $\sigma_0$ and two sets of parameters: $\{\tilde{a}\longrightarrow1,~ X\longrightarrow1,~ \xi\longrightarrow0\}$ (left panel) and $\{\tilde{a}\longrightarrow1,~ X=1.2,~ \xi=\pi/12\}$ (right panel).}
		\label{Neg}
	\end{center}
\end{figure}

Throughout this paper, the hot plasma we are interested in
is relativistic, with polytropic index $\Gamma=4/3$. Then, the energy extraction from the BH via MR happens when \cite{Comisso:2020ykg}
\begin{equation}\label{conditionsenergy}
\epsilon^\infty_-<0\,  \quad    {\rm and}   \quad   \Delta \epsilon^\infty_+=\epsilon^\infty_+ - \left( {1-\frac{\Gamma}{\Gamma-1}  \varpi } \right) >0 \,.
\end{equation}
By putting the expressions $\alpha,~v_K$, and $\beta^\phi$ in (\ref{energuisMagnet}), the final form of $\epsilon^\infty_\pm$ becomes complicated and thereby providing an analytical investigation of condition (\ref{conditionsenergy}), which does not seems to be straightforward.
However, in Fig. \ref{Neg} we  check the condition (\ref{conditionsenergy}) for a given set of involved parameters.
The left panel clearly shows that in the presence of LSB parameter $l$,
one can optimally extract energy even with  values of $\sigma_0$ below $1/3$, in addition to $X\longrightarrow1,~ \xi\longrightarrow0$.
This is importance in the sense that Comisso and Asenjo \cite{Comisso:2020ykg} have already shown that, for the optimal energy extraction from the standard extremal Kerr BH under conditions $X\longrightarrow1,~ \xi\longrightarrow0$, it is required that $\sigma_0>1/3$. As a result, adding the LSB parameter in the background results in optimal energy extraction via MR with values smaller than a lower bound $\sigma_0>1/3$. By increasing $\sigma_0$, we have a wide range of $l$, which can satisfy the condition $\epsilon^\infty_-<0$.
The right panel includes this message that by increasing values $X$ and $\xi$, more negative values of the allowed range of $l$ can play a role in favor of energy extraction. Note that, depending on the large-scale magnetic field configuration, the value of $\xi$ can be close to $\pi/4$. However, there is no conclusive agreement on the value of orientation angle $\xi$, and the exact value of $\xi$ depends on multiple factors. It is interesting that the numerical simulations signal small values (as well as time-dependent ones) so the value is commonly is within the range $\xi\in [\pi/12,\pi/6]$; see, e.g., \cite{Bransgrove:2021heo,Semenov:2004ib}. Throughout this paper, we prefer to keep the magnetic field orientation $\xi$ at the lower limit of the range above.

In what follows,  by doing a parameter space analysis we show that in light of an additional parameter $l$, there will be more possibilities to satisfy  (\ref{conditionsenergy}) and subsequently affect energy extraction from the BH. So it is expected that the underlying framework to analyze rotational energy extraction from the BH via MR is richer than its standard counterpart.

\begin{figure}
\begin{center}
\includegraphics[scale=0.45]{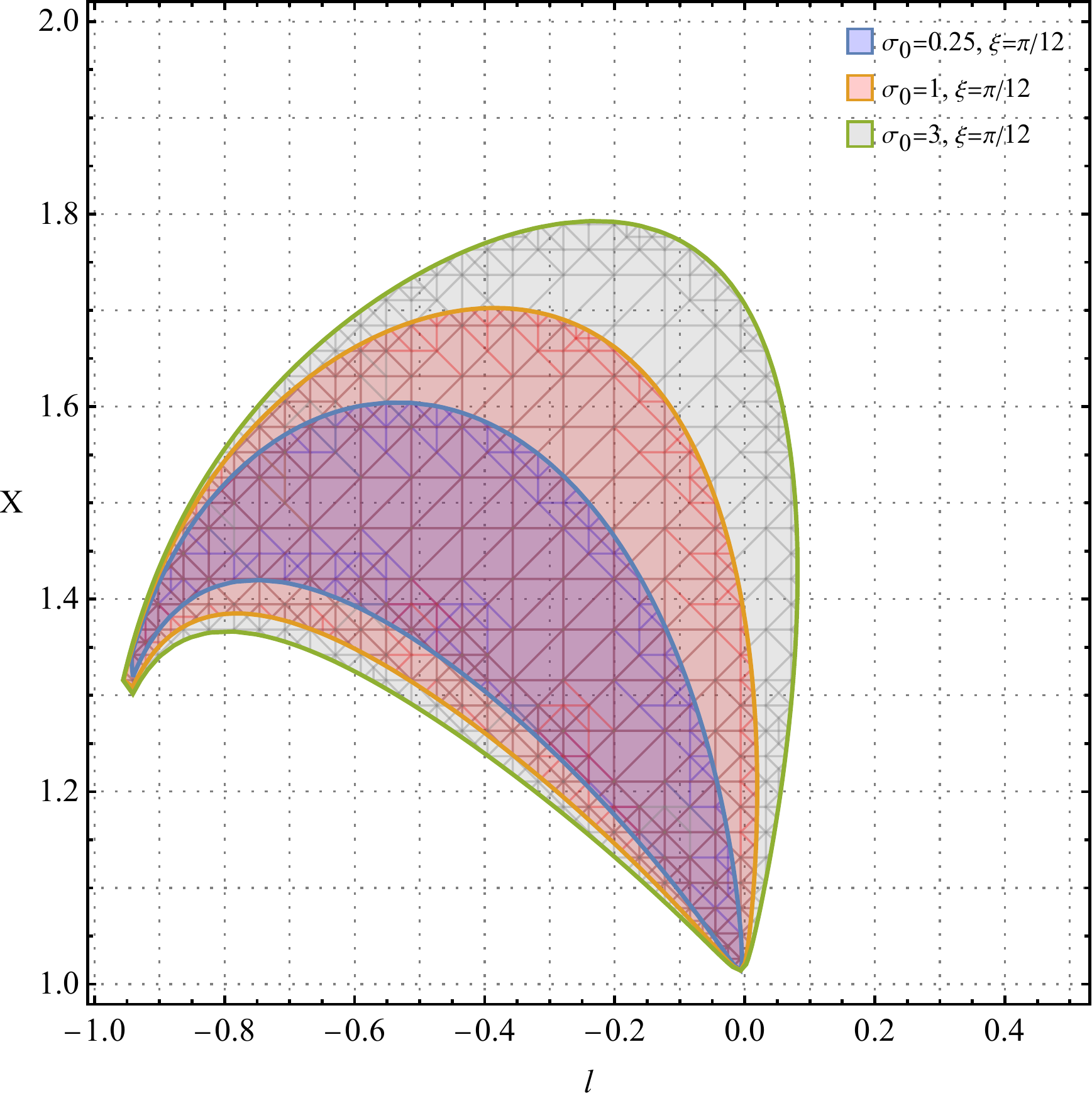}~~~
\includegraphics[scale=0.45]{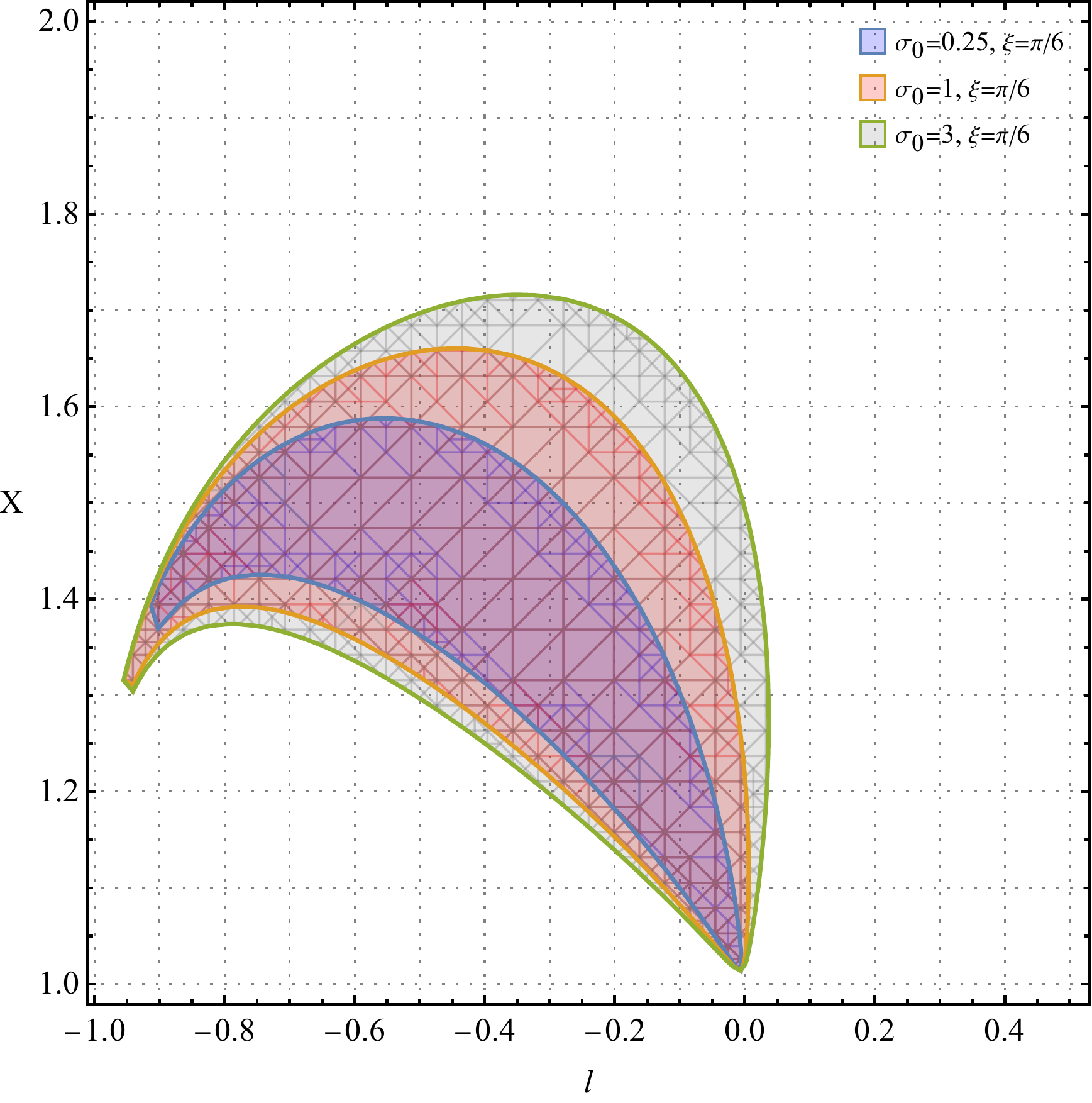}
\includegraphics[scale=0.45]{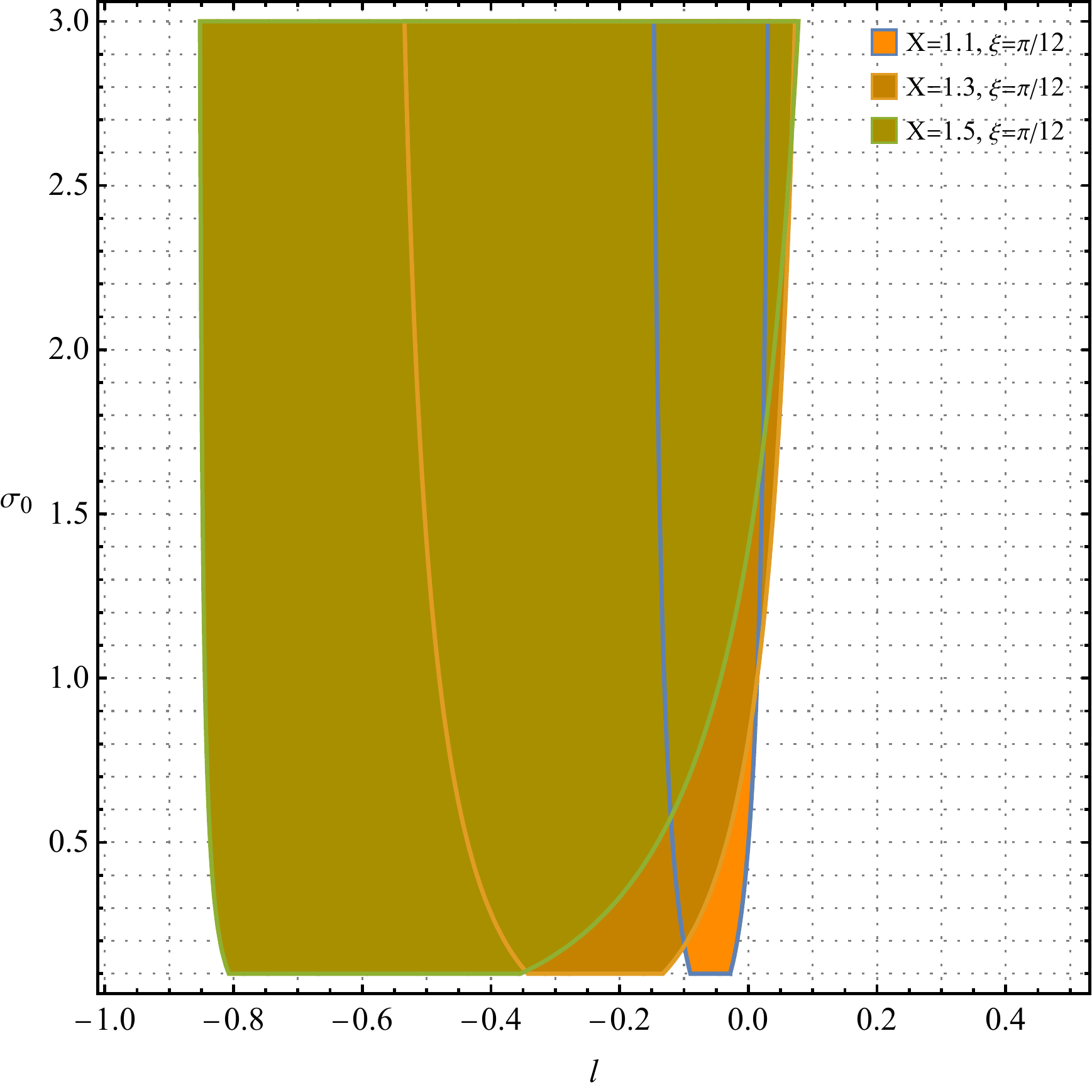}~~~
\includegraphics[scale=0.45]{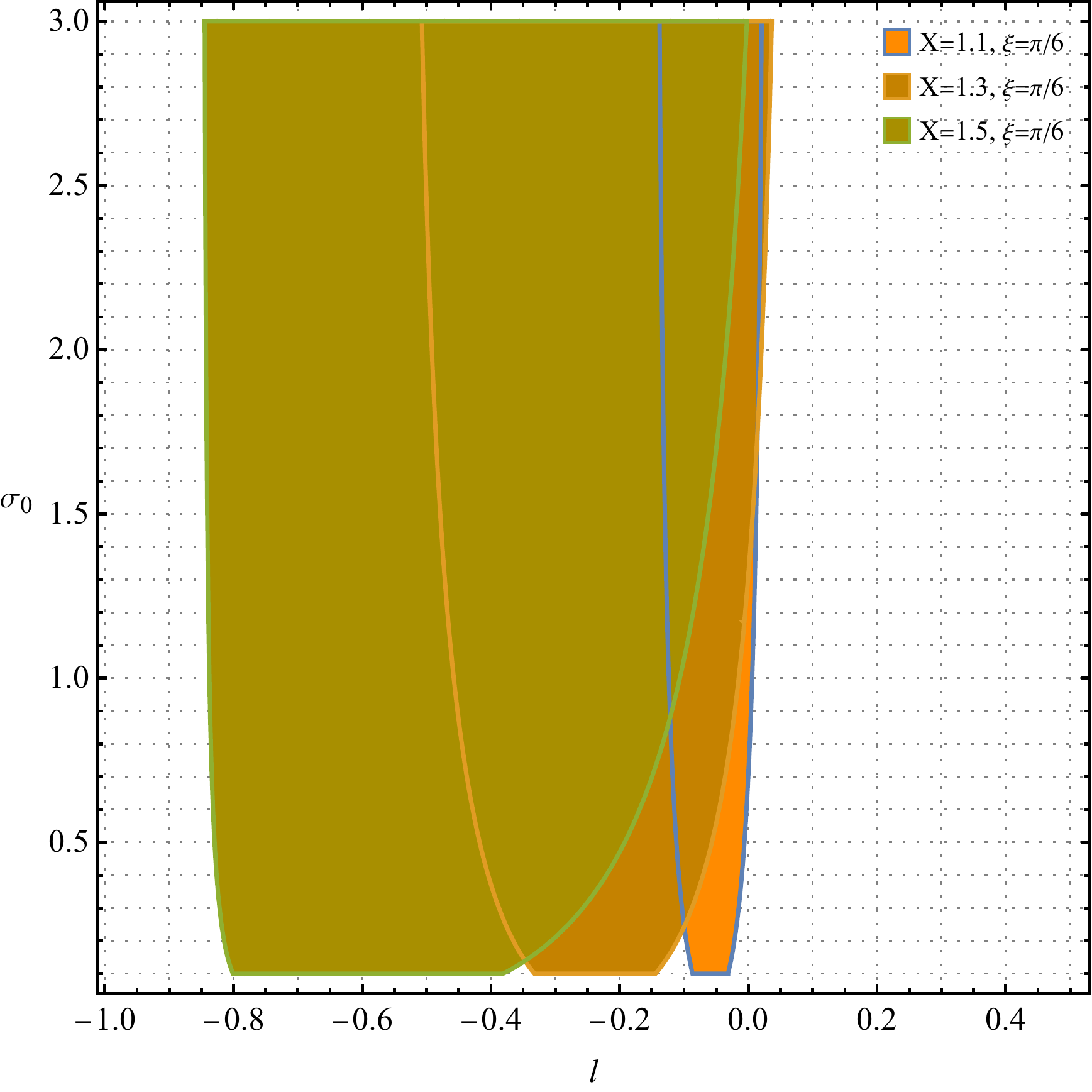}
\caption{\textbf{Top row:} regions of the phase space $(l,X)$ that meet the condition (\ref{conditionsenergy}). \textbf{Bottom row:} regions of the phase space $(l,\sigma_0)$ that meet the condition (\ref{conditionsenergy}).}
		\label{PS1}
	\end{center}
\end{figure}
\subsection{Analysis via parameter space}
By focusing on solutions of Eq. \eqref{energuisMagnet} and comparing them to the standard Kerr BH, we have an additional parameter $l$ indicating LSB in the background. Here we analyze the viability of energy extraction via MR in the interplay with this new background parameter. Particularly, in Fig. \ref{PS1} we display the regions of the phase spaces $(l,X)$ and $(l,\sigma_0)$ that satisfy conditions $\epsilon^\infty_- <0$ and $ \Delta \epsilon^\infty_+  >0$.
We use a MR process around a fast-spinning BH $(\tilde{a}=0.99)$ surrounded by the plasma with different values of the magnetization parameter $\sigma_0$ (top row of Fig. \ref{PS1}) and different values of the location of reconnection $X$ (bottom row of Fig. \ref{PS1}), which have the orientation angles $\xi=\pi/12,~\pi/6$. 

In these figures we show that, unlike the standard Kerr BH in the presence of negative values for Lorentz-violating parameter $l<0$, plasma with negligible magnetization ($\sigma_0<1/3$) also has a chance of satisfying relevant conditions for energy extraction from the BH via MR. As one can see in Fig. \ref{PS1} (top row), by growing the magnetization of the plasma,  the region of the phase-space $(l, X)$ for the BH rotational energy extraction becomes wider. Of course, the area allowed for the phase space $(l, X)$ depends on the orientation angle $\xi$, as one can see in the left and right panels. The bottom row in Fig. \ref{PS1} in coordination with the top row, openly shows that, as the location of MR (X points) moves away from the inner boundary of the ergosphere $(\sim1)$ towards the outer boundary $(\sim2)$, more values of $l<0$, play a role in energy extraction. In total, we see that the positive LSB parameter $l>0$ virtually does not play a role in the energy extraction from BH via the MR mechanism.

%%%%%%%%%%%%%%%%%%%%%%%%%%%%%%%%%%%%%%%%%%%%%%%%%%%%%%%%%%%%%%%%%%%%
\section{power and efficiency in the presence of Lorentz invariant violation}\label{power}

In this section, we evaluate the rate of energy extraction induced by the MR process from a fast-spinning background in which Lorentz symmetry is broken. In essence, this depends on the amount of plasma with negative energy at infinity that is swallowed by the BH in the unit time, meaning that, to have a high energy extraction rate, a MR with a high rate is essential. However, the BH in question has a Lorentz-violating parameter $l$ which is expected to affect this relation. The power $P_{\rm extr}$ per unite of enthalpy extracted from the BH by the escaping plasma can be estimated as \cite{Comisso:2020ykg}
\begin{equation} \label{Pextr}
\mathcal{P}_{\rm extr} = \frac{P_{\rm extr}}{\mathscr{H}}=- \epsilon_-^\infty A_{\rm in} U_{\rm in}  \, ,
\end{equation}
where $U_{\rm in} = \mathcal{O}(10^{-1})$ and $\mathcal{O}(10^{-2})$ \cite{Comisso:2016ima} for the collisionless and collisional regimes, respectively. In the above equation $A_{\rm in}$ is the cross-sectional area of the inflowing plasma, which for high spinning BHs, can be estimated as ${A}_{\rm in} \sim (r_{out}^2 - r_{{\rm ph}}^2)$. 

\begin{figure}
	\begin{center}
		\includegraphics[scale=0.42]{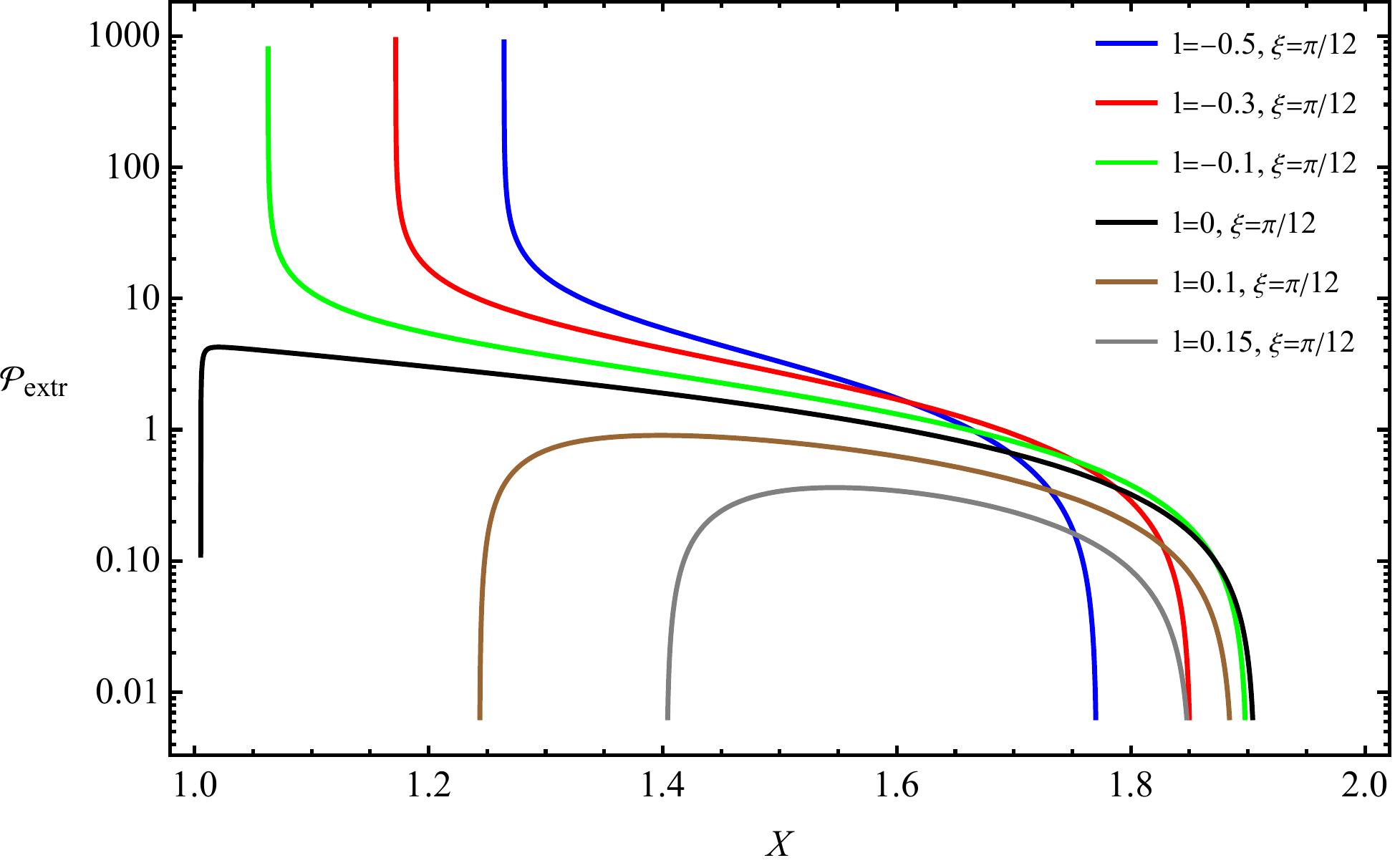}~~
		\includegraphics[scale=0.42]{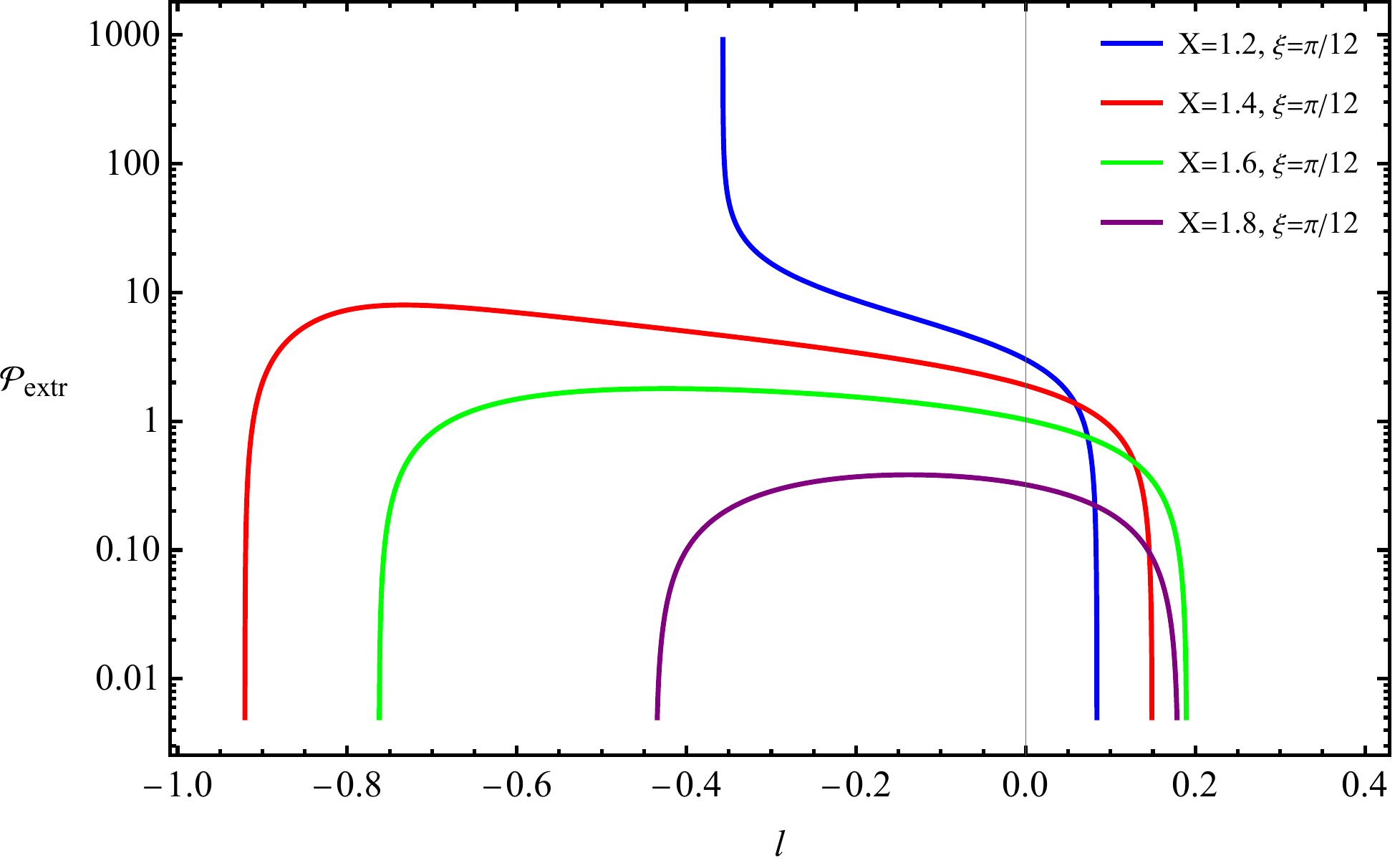}
		\caption{\textbf{Left:} log-plot of power per unit of enthalpy $\mathcal{P}_{\rm extr}$ as a function of the $X$-point location for a rapidly spinning BH ($\tilde{a} = 0.99$) and different values of Lorentz-violating parameter $l$. \textbf{Right:} Power per unit of enthalpy $\mathcal{P}_{\rm extr}$ as a function of the Lorentz-violating parameter $l$ for a rapidly spinning BH ($\tilde{a}= 0.99$) with different values of the $X$-point locations. In both panels, we have also set $M=1$ and $\sigma_0 = 10$.}
		\label{powe}
	\end{center}
\end{figure}
\begin{figure}
\begin{center}
		\includegraphics[scale=0.4]{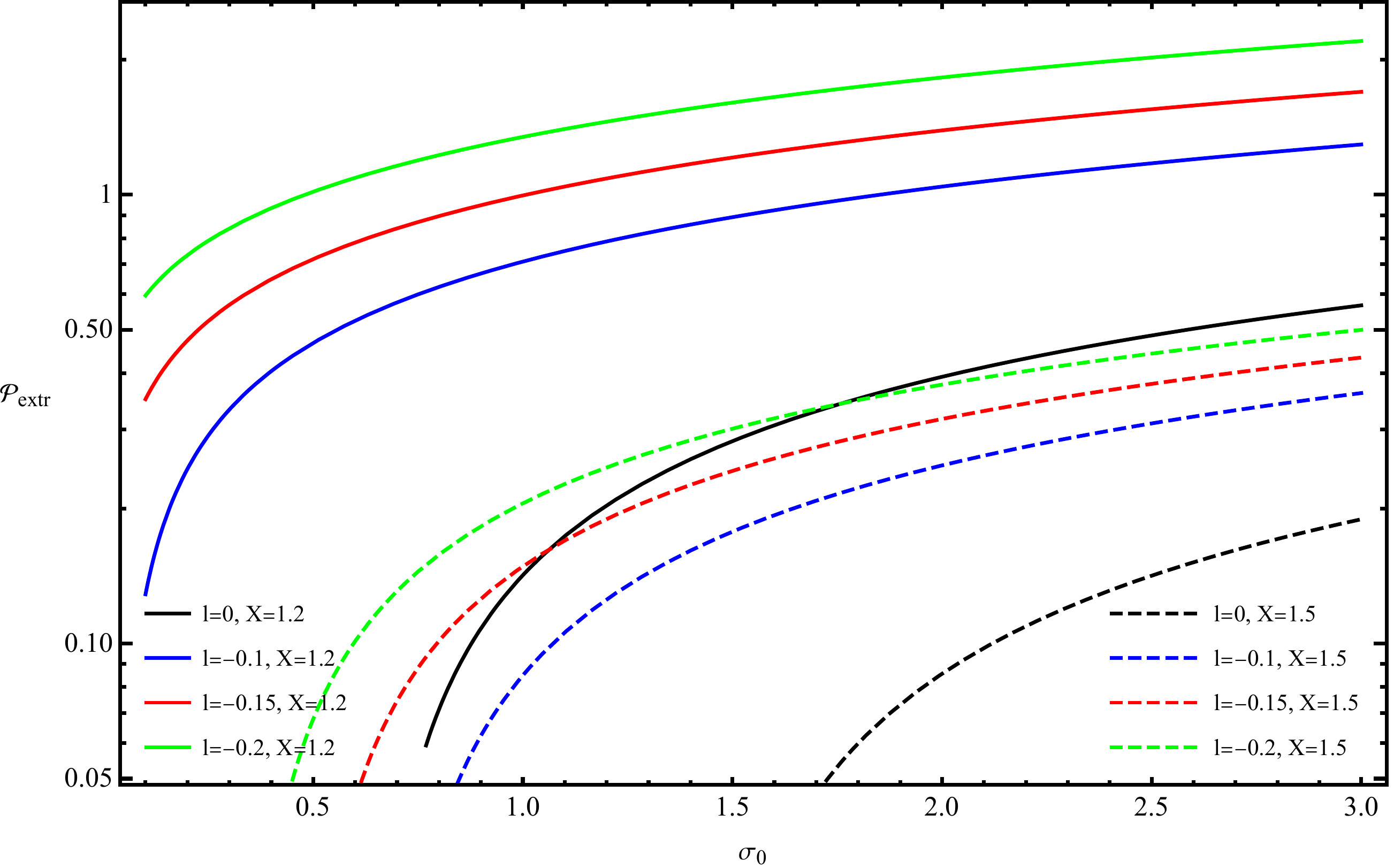}~~~
		\caption{Log-plot of power per unit of enthalpy $\mathcal{P}_{\rm extr}$ as a function of the plasma magnetization parameter $\sigma_0$ for a rapidly spinning BH with spin parameter $\tilde{a} = 0.99$ and different values of the Lorentz-violating parameter $l$ and $X$-point location. Here, we have fixed numerical value $\xi=\pi/12$. }
		\label{power1}
	\end{center}
\end{figure}
In Fig. \ref{powe}, by adopting the collisionless approximation, i.e., $U_{\rm in} = 0.1$, we respectively show the power per enthalpy $\mathcal{P}_{\rm extr}$ as a function of the $X$-point location of reconnection and the Lorentz-violating parameter $l$ for a rapidly rotating BH. The left panel shows that the negative values $l<0$ extract more power from the BH, compared to $l\geq0$. By increasing negative values of the Lorentz-violating parameter, the peak of power forms farther from $X\sim1$ in $X$-point locations. In the right panel, we see that $l<0$, along with the $X$-point location close to $X\sim1$, results in the most power from the BH, so by approaching $X\sim2$, it drops. In Fig. \ref{power1}, clearly we see that by turning on $l<0$ in the background, for the plasma magnetization parameter below the case of $l=0$, there are power curves that indicate the possibility of energy extraction. So, in the presence of $l<0$, if the MR occurs around $X\sim1$, then one can expect the energy extraction from the fast-rotating BH enclosed by the plasma with weak magnetization. Note that the general trend of curves in Figs. \ref{powe} and \ref{power1} is independent of the fixed value for the orientation angle $\xi$.
 
\begin{figure}
	\begin{center}
		\includegraphics[scale=0.5]	{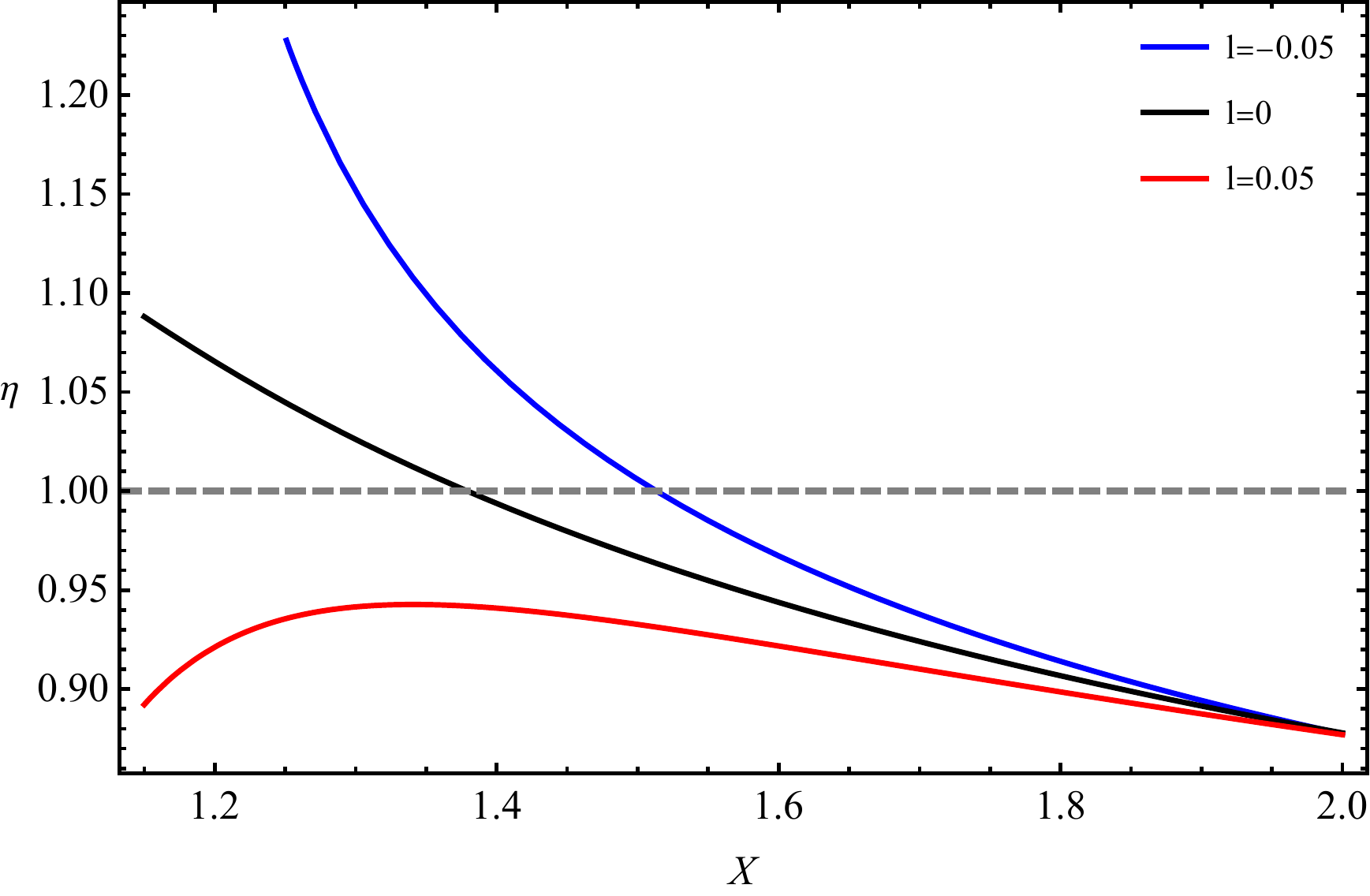}~~~
		\includegraphics[scale=0.5]	{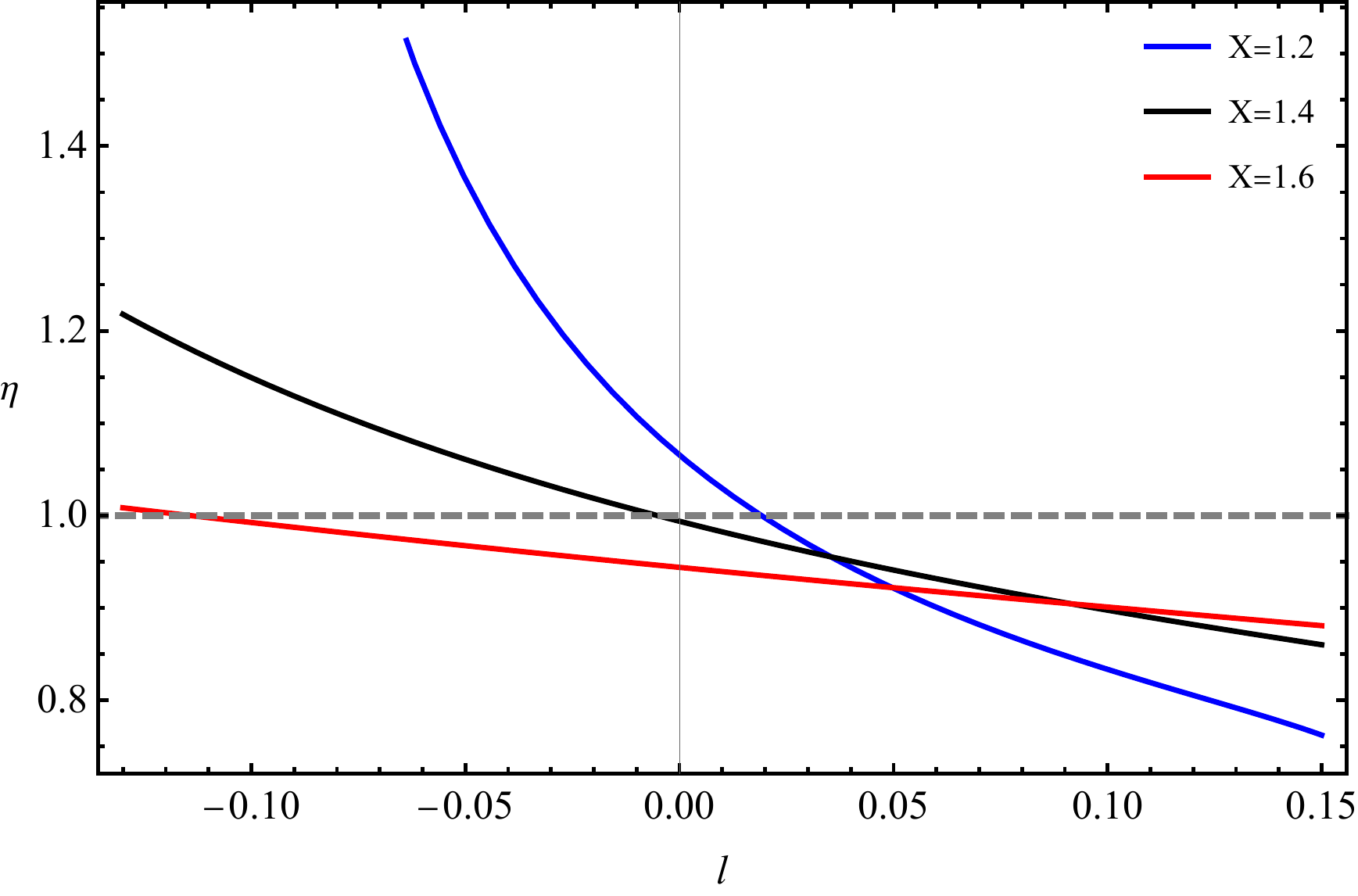}
		\caption{\textbf{Left:} efficiency $\eta$ of the reconnection mechanism as a function of the dominant $X$-point location with different values of the Lorentz-violating parameter $l$. \textbf{Right:} efficiency $\eta$ of the reconnection process as a function of the Lorentz-violating parameter $l$ with different values of the dominant $X$-point location. For both panels, we use a rapidly rotating BH with spin parameter $\tilde{a}=0.99$, which is surrounded by a plasma with magnetization $\sigma_0 = 1$ and orientation angle $\xi=\pi/12$. The horizontal gray-dashed line shows $\eta=1$.}
		\label{EF}
	\end{center}
\end{figure}

\begin{figure}
	\begin{center}
		\includegraphics[scale=0.5]{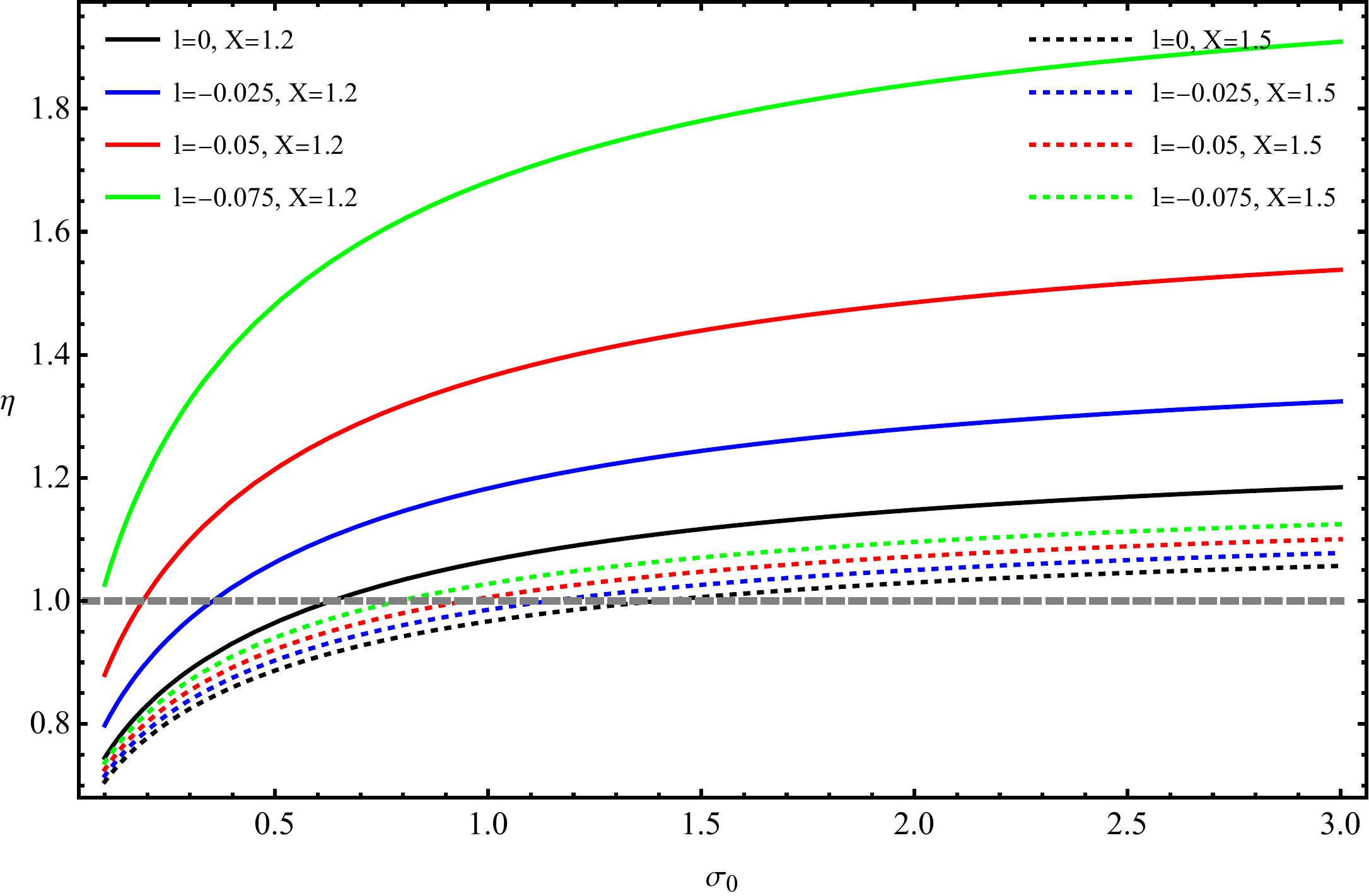}~~~
		\caption{Efficiency $\eta$ of the reconnection mechanism as a function of the plasma magnetization $\sigma_0$ for different values of Lorentz-violating parameter $l$ and $X$-point location. Here, we have fixed numerical values $\tilde{a}=0.99$ and $\xi=\pi/12$. The horizontal gray-dashed line shows $\eta=1$.}
		\label{EF1}
	\end{center}
\end{figure}
Although the mechanism in question for energy extraction via MR generates energetic plasma outflows that steal energy from the BH, it needs magnetic field energy to operate. 
Actually, the role of magnetic energy is to redistribute the angular momentum of the particles in such a way as to produce particles with negative energy (from the viewpoint of the infinity observer) and particles escaping to infinity. Conventionally, the efficiency of the plasma energization process via MR is defined as follows \cite{Comisso:2020ykg}:
\begin{equation}  \label{eff}
	\eta =  \frac{\epsilon^\infty_+}{\epsilon^\infty_+ + \epsilon^\infty_-}   \,  ,
\end{equation}
where in the case of $\eta > 1$, rotational energy extraction from the BH occurs. Figure \ref{EF} shows the efficiency $\eta$ as a function of the $X$-point location and LSB parameter $l$
for a reconnection layer with the plasma magnetization $\sigma_0=1$. Here, also, we can see that the negative values $l<0$ compared to $l\geq0$ result in more efficiency of the reconnection process, which has phenomenological worth. The efficiency $\eta$ increases for reconnection $X$ points that are closer to the BH event horizon and drops below unity for farther reconnection $X$ points, as is clear from the right panel.
The role of increasing efficiency by plasma magnetization $\sigma_0$ is also shown in Fig. \ref{EF1}. Here, in the presence of $l<0$, for the plasma with weak magnetization below what is expected from $l=0$, the efficiency becomes bigger than unity, $\eta>1$. Of course, it is more efficient if the MR mechanism occurs around $X\sim1$ in a plasma with strong magnetization $\sigma_0$.
  Overall, the results obtained here and also in Figs. \ref{powe} and \ref{power1} are in agreement with Fig. \ref{Neg}.

\begin{figure}
	\begin{center}
		\includegraphics[scale=0.35]{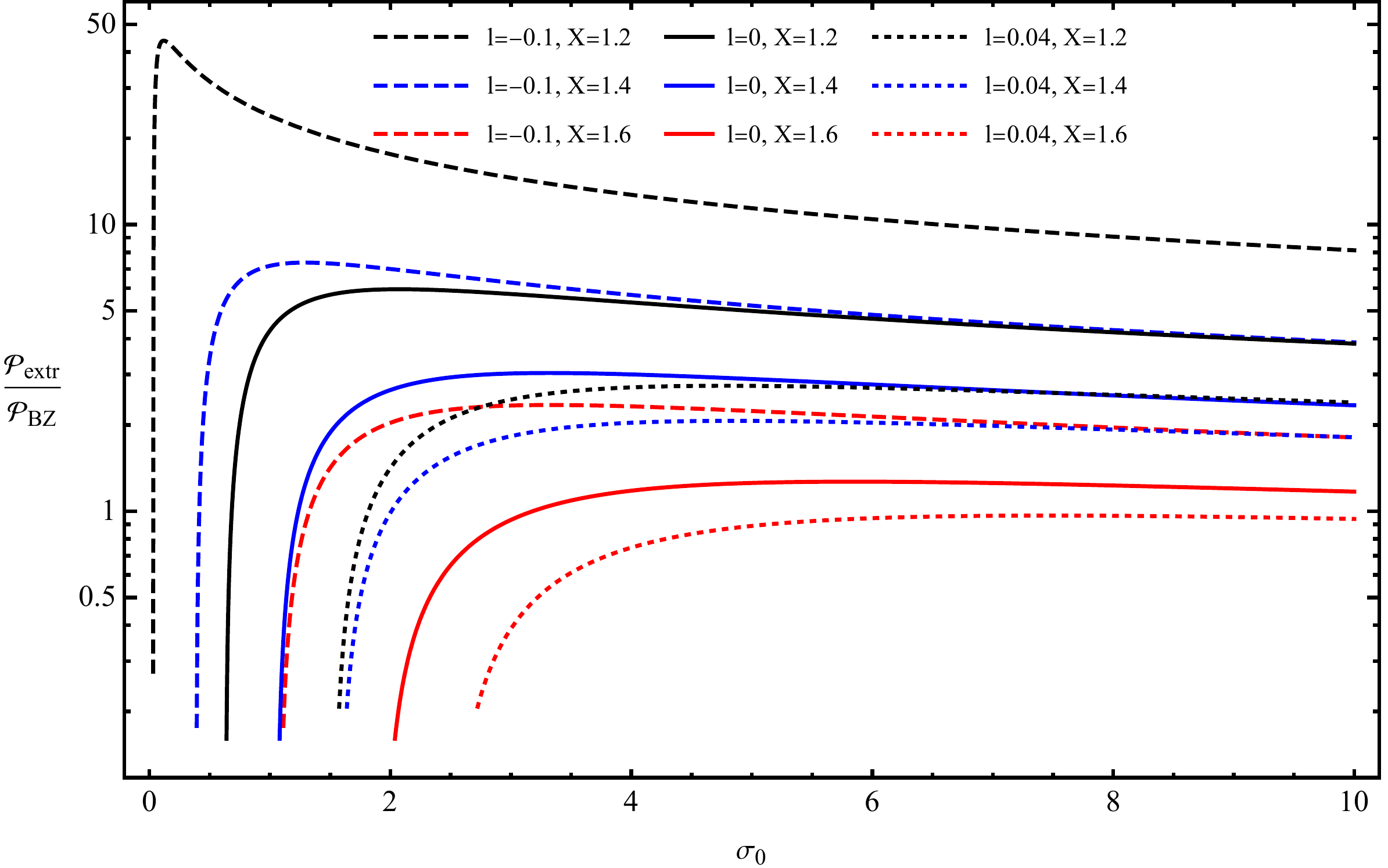}~~~
		\includegraphics[scale=0.35]{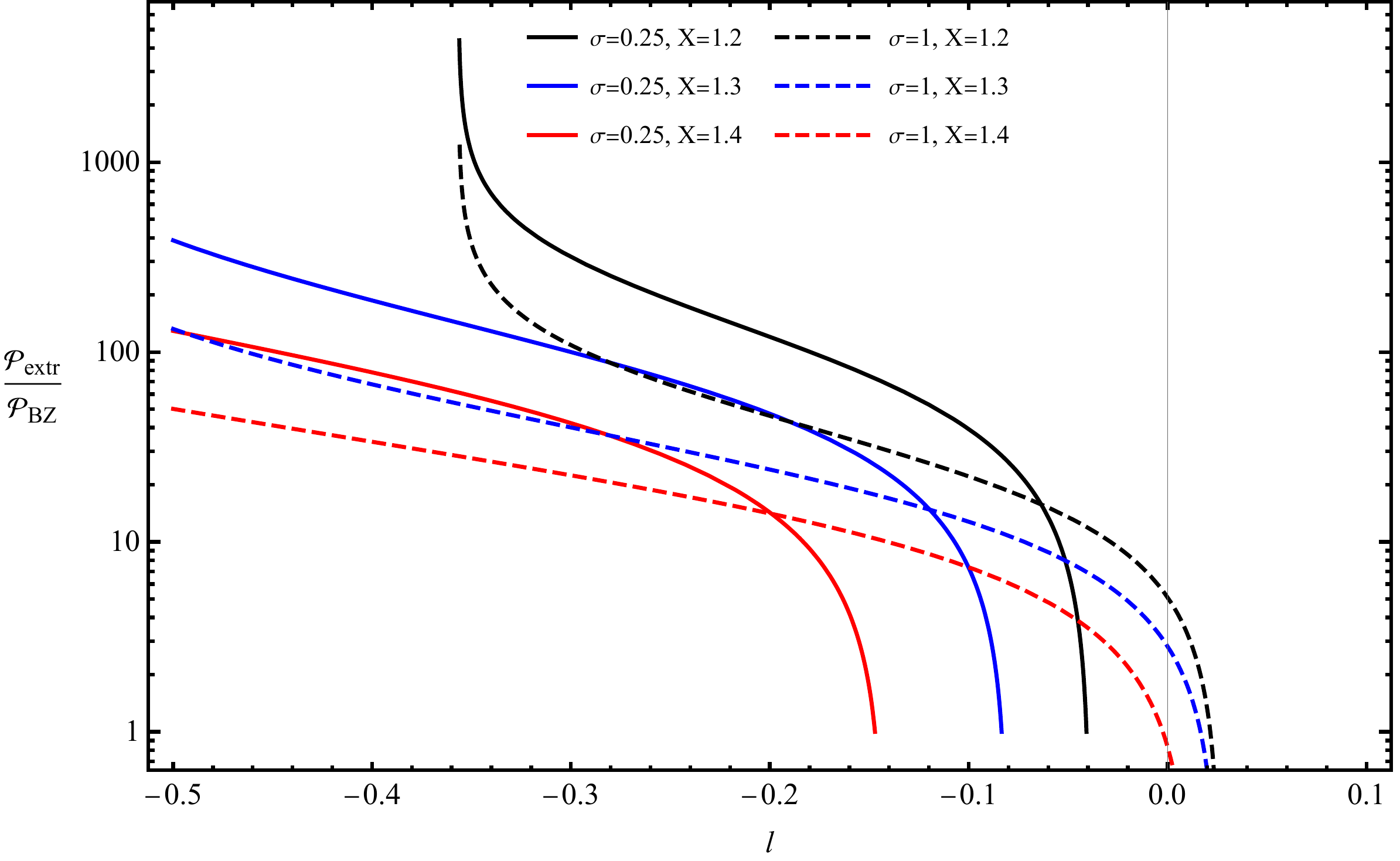}
		\caption{\textbf{Left:} log-plot of the power ratio $\frac{P_{\rm extr}}{P_{\rm BZ}}$ in terms of small values of plasma magnetization $\sigma_{0}$ for different values of the Lorentz- violating parameter $l$ and $X$-point location of reconnection. \textbf{Right:} Log-plot of the power ratio $\frac{P_{\rm extr}}{P_{\rm BZ}}$ in terms of the Lorentz-violating parameter $l$ for different values of plasma magnetization $\sigma_{0}$ and $X$-point location of reconnection. We have fixed values
			$\tilde{a}=0.99,~\xi=\pi/12$ along with the numerical coefficients $\kappa \approx 0.044,~c_1 \approx 1.38,~c_2 \approx -9.2$, from \cite{Tchekhovskoy:2009ba}.}
		\label{Com}
	\end{center}
\end{figure}

\begin{figure}
\begin{center}
	\includegraphics[scale=0.5]{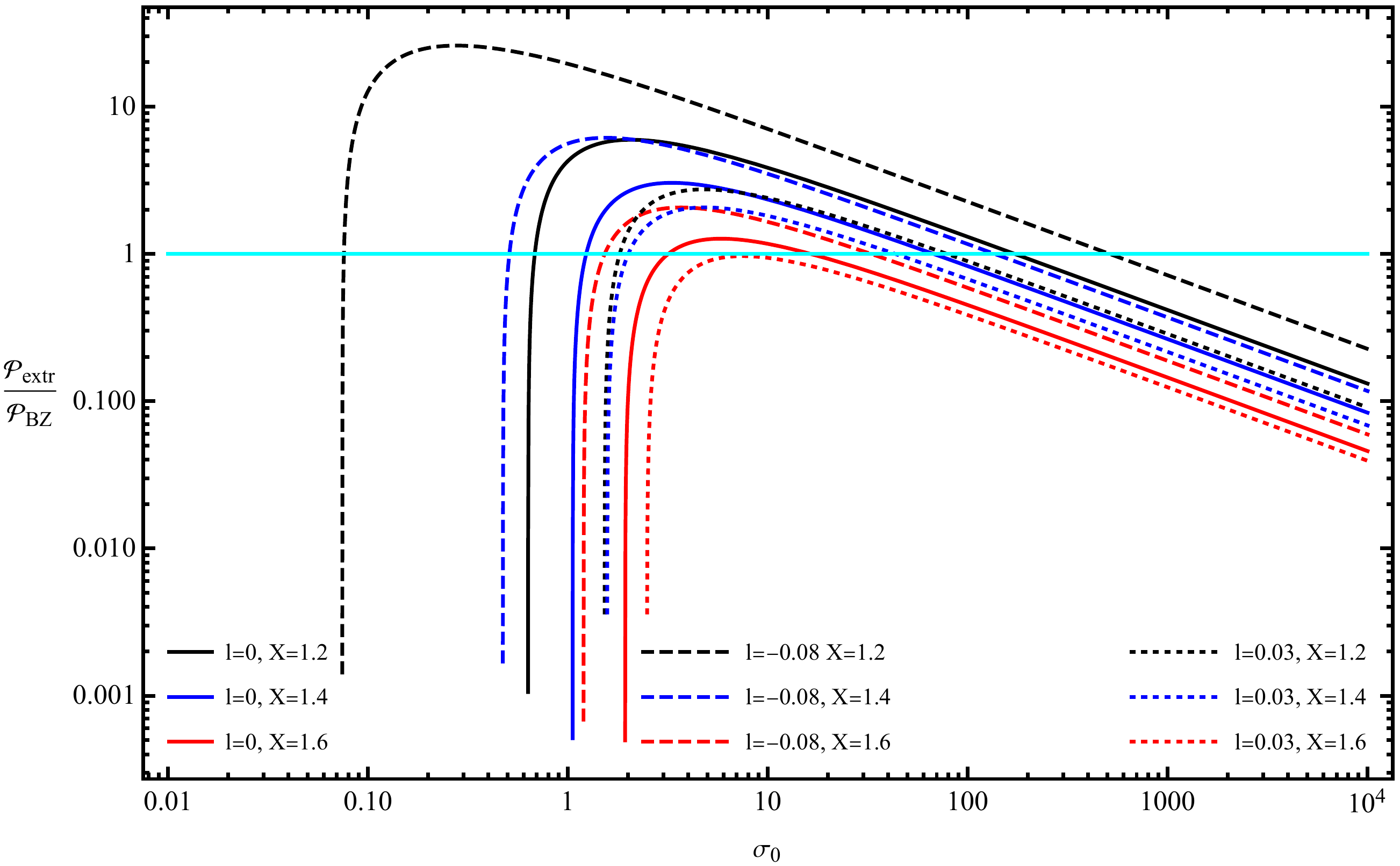}~~~
		\caption{Log-log plot of the power ratio $\frac{P_{\rm extr}}{P_{\rm BZ}}$ in terms of large values of the plasma magnetization $\sigma_{0}$ for different values of the Lorentz- violating parameter $l$ and $X$-point location of reconnection. The cyan horizontal line represents the boundary $P_{extr}/P_{BZ}=1$. }
		\label{EL}
	\end{center}
\end{figure}
\subsection{Comparing with energy extraction via BZ mechanism}
To obtain a clear intuition of the capability to extract energy in the presence of the LSB parameter $l$,  it is a good idea to compare the power arising from the MR and  BZ mechanisms. Recall that, in the BZ  mechanism, the rotational energy is extracted electromagnetically by the magnetic field sourced by the material accreting around the BH. 
It would be interesting to note that due to the lack of the change of connectivity of the magnetic field lines in a force-free magnetosphere surrounding the BH \cite{Blandford:1977ds}, the MR does not occur in the framework of the BZ process \cite{Comisso:2019hem}. 
The BZ mechanism, in essence, relies on the magnetic flux $\Phi_{BH}$ threading the BH, and for this reason, the plasma inertia is negligible. In other words, in the case of enclosing a spinning BH by an external magnetic field, according to Lorentz transformation, an electric field appears in the corotating frame, which induces separation of charges, meaning an electric current in the inertial frame \cite{Konoplya:2021qll}. In this way, the rotational energy of the BH is transferred into the energy of the currents outside the BH. However, the energy extraction mechanism proposed by Comisso-Asenjo relied on the plasma having finite inertia, along with the magnetic field being significant only for the acceleration of the plasma in opposite directions. Therefore, the two mechanisms are distinct, and they extract energy from the BH in two distinctive ways.

The rate of BH energy extraction via the BZ mechanism up to leading order of the angular velocity of the event horizon $\Omega_H$ is given by \cite{Tchekhovskoy:2009ba}
\begin{equation} \label{P_BZ}
P_{\rm BZ} = \dfrac{\kappa}{16\pi} \Phi_{\rm BH}^2 \Omega_H^2+ \mathcal{O}(\Omega_H^4)\, ,
\end{equation} where $\Phi_{\rm BH} = 2\pi \int_{0}^\pi |B^r| \sqrt{-g}~d \theta$ ($g$ is the determinant of the metric tensor) refers to the magnetic flux threading the BH horizon and $\kappa$ is a numerical constant related to the magnetic field configuration \cite{Pei:2016kka}. Note that by including higher order terms of $\Omega_H$,
Eq. (\ref{P_BZ}) can be reexpress as 
\begin{equation} \label{P_BZR}
P_{\rm BZ} = \dfrac{\kappa}{16\pi} \Phi_{\rm BH}^2 \Omega_H^2
	(1+c_1\Omega_H^2+c_2\Omega_H^4+...)\, ,
\end{equation} where $c_{1,2,...}$ denote certain numerical coefficients. By taking the elements of the metric tensor (\ref{elemen}), we have $ \Phi_{\rm BH}=4\pi\sqrt{l+1}|B_0| r_H^2$ and $\Omega_H=(-\frac{g_{t\phi}}{g_{\phi\phi}})_{r_{H}}=\frac{\tilde{a}}{2r_{H}}$.
Now by inserting these into  Eq. (\ref{P_BZR}) and also using Eq. (\ref{Pextr}),  the ratio $\frac{P_{\rm extr}}{P_{\rm BZ}}$ reads 
\begin{equation} \label{powerratiowithBZ1}
\frac{\mathcal{P}_{\rm extr}}{\mathcal{P}_{\rm BZ}} =\frac{ -4  U_{\rm in} (r_{out}^2 - r_{{\rm ph}}^2)~ \epsilon_-^\infty} {\kappa \pi(l+1)\sigma_0 \tilde{a}^2 r_H^2\bigg(1+\frac{c_1\tilde{a}^2}{4r_H^2}+\frac{c_2\tilde{a}^4}{16r_H^4}+...\bigg)}\,,~~~~~~\mathcal{P}_{\rm BZ}=\frac{P_{\rm BZ}}{\mathscr{H}}.
\end{equation}
An important point to note about the above equation is that it is nothing 
but a very simple estimate for the magnetic flux $\Phi_{BH}$ threading the BH horizon. A precise evaluation of $\Phi_{BH}$ can only be performed with \textit{ab initio} numerical simulations. Without taking the relevant simulations, it can provide just an order-of-magnitude estimate, indicating a qualitative comparison between the proposed mechanism and the popular BZ mechanism.
The left and right panels in Fig. \ref{Com} respectively show the ratio $\frac{P_{\rm extr}}{P_{\rm BZ}}$ given by the Eq. \eqref{powerratiowithBZ1} as a function of the plasma magnetization $\sigma_0$ and Lorentz-violating parameter $l$ for different values of other involved parameters. As can be seen, modification arising from the LSB in the background affects the ratio $\frac{P_{\rm extr}}{P_{\rm BZ}}$ in comparison with its standard counterpart, so that ratios for $l<0$ and $l>0$ are respectively bigger and smaller than $l=0$. This means that embedding a negative LSB parameter $l<0$ into the background of a Kerr-like BH increases the power of energy extraction via MR compared to the BZ mechanism. In other words, if the LSB parameter happens to be negative, then the MR is a more efficient mechanism relative to the BZ mechanism for energy extraction. Similar to the standard MR mechanism, at sufficiently large plasma magnetization $\sigma_0$, the Lorentz-violating solution at hand also drops compared to the BZ one because the force-free electrodynamics (BZ) based solution dominates as the plasma magnetization increases.
In other words, this happens in the form of a transition from $P_{extr}/P_{BZ} > 1$ to $P_{extr}/P_{BZ} < 1$ at a certain threshold value of $\sigma_0$ in which MR becomes subdominant to the BZ process; see Fig. \ref{EL}. However, as is evident, the LSB parameter affects the threshold value of $\sigma_0$, so it is bigger and smaller for $l<0$ and $l>0$, respectively.

%%%%%%%%%%%%%%%%%%%%%%%%%%%%%%%%%%%%%%%%%%%%%%%%%%%%%%%%%%%%
\section{Conclusion}\label{Con}
Magnetic reconnection is a fundamental trait of astrophysical and laboratory plasmas; it occurs under some conditions and suddenly releases magnetic energy.
Fast MR is one of the energy extraction mechanisms sourced by oppositely directed magnetic field lines (arising from frame-dragging of a spinning BH) close to the equatorial plane, as proposed in the model by Comisso and Asenjo \cite{Comisso:2020ykg} outlined in Fig. \ref{bh}. Once a spinning BH in an external magnetic field is enclosed, the reconnection of antiparallel magnetic field lines within the region of the ergosphere causes the falling particles into the BH event horizon to have negative energy at infinity.
On the other hand, the particles with positive energy at infinity escape with more energy than before, meaning they they obtained energy from the BH. To be more specific, the MR accelerates the plasma in opposite directions one part in the direction of the BH rotation and another one in the opposite direction falling into the BH. So, BH energy extraction happens provided that the plasma that has fallen into the BH from the perspective of the infinity observer has negative energy, meaning that the plasma accelerating toward infinity obtained obtained energy from the BH. 

In this paper, we have extended the analysis on the extraction of BH rotational energy through fast MR for the case of a Kerr-like BH solution modified by the LSB parameter $l$, which comes from the non-zero VEV of the background bumblebee field. We first derived an analytical expression for the energy at infinity associated with the accelerated or decelerated plasma, which is a function of the critical parameters ($\tilde{a}$, $X$, $l$, $\sigma_0$, $\xi$). We have shown that, unlike the standard Kerr, in the presence of the LSB parameter $l<0$, energy extraction from a fast-spinning BH enclosed by plasma with negligible magnetization below the lower bound derived in \cite{Comisso:2020ykg} is possible.
By conducting this analysis within phase spaces $(l,X)$ and $(l,\sigma_0)$, we found that in the presence of $l<0$, plasma with negligible magnetization also has a chance of satisfying relevant conditions for energy extraction from BH via MR. It occurs at distances near the inner boundary of the ergosphere, i.e., event horizon. 

We evaluated the role of the LSB parameter $l$ on the power of energy extraction and efficiency of the plasma energization process through MR. Our analysis explicitly showed that $l<0$ has an amplifying role in both quantities, which is significant from a phenomenological viewpoint.  In the end, we compared the power of energy extraction from the bumblebee-based BH via a fast MR process with the one that can be extracted via the BZ mechanism. We have shown that by embedding negative LSB parameter $l<0$ into the background of a Kerr-like BH, we can increase the power of energy extraction via MR enhancement considerably compared to what is expected of the BZ mechanism for this background.

To summarize, energy extraction from the BH via the reconnection process in the rotating BH with the Lorentz symmetry broken by $l<0$ is more efficient than the standard case addressed in \cite{Comisso:2020ykg}. A similar result previously has released in \cite{Khodadi:2021owg} for superradiance as another well known energy extraction mechanism for BH.

As an idea for the future, one can investigate the role of electric charge on the rate of energy extraction via MR mechanism and its competition with the contributions from the LSB parameter. For instance, the role of electric charge on the BZ mechanism has been investigated recently in \cite{King:2021jlb,Komissarov:2021vks}.
Theoretically, considering the contribution of electric charge is well motivated from several aspects: first, according to the no-hair theorem, any astrophysical BH described by three observable parameters- mass, spin, and electric charge \cite{Khodadi:2021gbc,Rahmani:2020vvv}. Second, unlike the argument that the presence of plasmas around astrophysical BHs leads to a prompt discharge, the relevant environment is prone to some processes at both the classical and relativistic levels that result in the production of a small nonzero electric charge \cite{Zajacek:2019kla}. Third, in a recent scenario related to Wald's charge \cite{Wald:1974np}, the authors have shown that a spinning BH boosted by a uniform magnetic field is prone to acquire an electric charge \cite{Adari:2021qmx}. 

\section{Acknowledgments}
The author is grateful to Luca Comisso for a critical reading of the manuscript and enlightening discussions and comments.

\end{document}